\documentclass[10pt,journal,compsoc,twoside]{IEEEtran}
\IEEEoverridecommandlockouts

\AtBeginDocument{%
  \providecommand\BibTeX{{%
    Bib\TeX}}}
    
\usepackage{url}
\usepackage{cite} 
\usepackage{amsmath,amssymb,amsfonts}
\usepackage{accents}
\usepackage[linesnumbered,ruled]{algorithm2e}
\usepackage[noend]{algpseudocode}
\usepackage{graphicx}
\usepackage{textcomp}
\usepackage[table,xcdraw]{xcolor}
\usepackage{xcolor}
\usepackage{comment}
\usepackage{multicol,multirow}
\usepackage[T1]{fontenc}
\def\BibTeX{{\rm B\kern-.05em{\sc i\kern-.025em b}\kern-.08em
    T\kern-.1667em\lower.7ex\hbox{E}\kern-.125emX}}
\usepackage{todonotes}
\usepackage{caption}
\usepackage{wrapfig}
\usepackage{subfig}
\newsavebox{\measurebox}
\usepackage{threeparttable}
\usepackage[font=small]{caption} 
\usepackage{enumitem}
\usepackage{tikz}

\usepackage[hidelinks]{hyperref}
\hypersetup{
    colorlinks=false, 
    linkcolor=blue,
    filecolor=magenta,      
    urlcolor=cyan,
    }

\makeatletter
\newcommand{\algorithmfootnote}[2][\footnotesize]{%
  \let\old@algocf@finish\@algocf@finish
  \def\@algocf@finish{\old@algocf@finish
    \leavevmode\rlap{\begin{minipage}{\linewidth}
    #1#2
    \end{minipage}}%
  }%
}
\makeatother

\newcommand*{\Scale}[2][4]{\scalebox{#1}{$#2$}}%

\begin{document}
\bstctlcite{IEEEexample:BSTcontrol}


\title{KnowSafe: Combined Knowledge and \\Data Driven Hazard Mitigation in \\Artificial Pancreas Systems} 
\author{Xugui~Zhou,~\IEEEmembership{Member,~IEEE,}
        Maxfield Kouzel,~\IEEEmembership{Member,~IEEE,}\\
        Chloe Smith,~\IEEEmembership{Member,~IEEE,}
        and~Homa~Alemzadeh,~\IEEEmembership{Member,~IEEE}
\IEEEcompsocitemizethanks{
\IEEEcompsocthanksitem Xugui Zhou is with Louisiana State University, Baton Rouge, LA 70803; Maxfield Kouzel, Chloe Smith, and Homa Alemzadeh are with the University of Virginia, Charlottesville, VA 22904 USA. \protect 

}


\thanks{Manuscript received November 2023, revised February 2025, May 2025, accepted June 2025. This work was supported in part by the Commonwealth of Virginia under Grant CoVA CCI: CQ122-WM-02 and by the National Science Foundation (NSF) under grants CNS-2146295 and CCF-2402941. (Corresponding author: Xugui Zhou. E-mail: xuguizhou@lsu.edu)}}

\markboth{IEEE TRANSACTIONS ON DEPENDABLE AND SECURE COMPUTING,~Vol. 0, No. 0, 2025}%
{ZHOU et al.: KnowSafe: Combined Knowledge and Data Driven Hazard Mitigation in Artificial Pancreas Systems}

\IEEEtitleabstractindextext{%
\begin{abstract} 
Significant progress has been made in anomaly detection and run-time monitoring to improve the safety and security of cyber-physical systems (CPS). However, less attention has been paid to hazard mitigation. This paper proposes a combined knowledge and data driven approach, \textit{KnowSafe}, for the design of safety engines that can predict and mitigate safety hazards resulting from safety-critical malicious attacks or accidental faults targeting a CPS controller. We integrate domain-specific knowledge of safety constraints and context-specific mitigation actions with machine learning (ML) techniques to estimate system trajectories in the far and near future, infer potential hazards, and generate optimal corrective actions to keep the system safe. Experimental evaluation on two realistic closed-loop testbeds for artificial pancreas systems (APS) and two real-world datasets for diabetes treatment demonstrates that \textit{KnowSafe} outperforms the state-of-the-art by achieving higher accuracy in predicting system state trajectories and potential hazards, a low false positive rate, and no false negatives. It also maintains the safe operation of the simulated APS despite faults or attacks without introducing any new hazards, with a hazard mitigation success rate of 92.8\%, which is at least 76\% higher than solely rule-based (50.9\%) and data-driven (52.7\%) methods. 

\end{abstract}

}

\maketitle
\IEEEdisplaynontitleabstractindextext

%
\IEEEpeerreviewmaketitle

\IEEEraisesectionheading{
\section{Introduction}
\label{sec:introduction}
}

%

\IEEEPARstart{C}{yber}-physical systems (CPS), {such as artificial pancreas systems (APS) for diabetes management, are} vulnerable to accidental faults or malicious attacks targeting their sensors, controllers, or actuators to cause catastrophic consequences  \cite{li2011hijacking, attack_pcs, zhou2022Strategic,zhou2022design}. 
Over the past decade, APS have been the subject of over 14 vulnerability reports \cite{CVEpump}, 92 U.S. Food and Drug Administration (FDA) recalls, and 2.73 million adverse event reports \cite{zhou2022design}. 
Despite advancements in safety and security through anomaly detection or run-time monitoring \cite{dsn2021zhou, SAVIOR2020,choi_ccs_18,ding2021mini}, effective hazard mitigation in CPS receives less attention and remains a significant challenge, especially for attacks or faults targeting CPS controllers (see Table 1). 

{Traditional approaches often} rely on redundant hardware or software sensors or components \cite{lyons1962use, Choi2020software} to maintain safe system operations after anomaly detection. However, these methods increase the system complexity and remain susceptible to common vulnerabilities shared across replicas.
{Other strategies involve} triggering a fail-safe mode (e.g., freezing insulin pumps in APS) \cite{zhang2011generic,paul2011review} or manual remediation (e.g., generating alerts, handing over control to the human operator \cite{zhou2022Strategic,meneghetti2019detection}), which may disrupt the normal system operation or struggle to ensure the safety of a manual recovery approach given the tight timing constraints and short reaction times. 

{Great efforts have also been made in designing} run-time monitors \cite{dsn2021zhou,desai2019soter} and simplex controllers \cite{Phan2017Simplex} that can act as a backup when the main controller is compromised. 
Model-based approaches in this area rely on developing linear or non-linear models of system dynamics and their interactions with the environment to detect anomalies \cite{attack_pcs} and generate recovery actions \cite{Linearrecovery2020Zhang,zhang2023real}. However, it is challenging to develop models that can fully capture complex system dynamics and unpredictable human physiology and behavior (e.g., glucose sensitivity to insulin in APS). Further, approximation errors between actual system states and model predictions can accumulate over time, limiting the effectiveness of these recovery approaches.

Data-driven approaches using machine learning (ML) have shown improved accuracy and success rate in attack detection and recovery~\cite{pidpiper2021, Choi2020software,luo_deep_2021}. Nevertheless, these approaches usually rely on black-box deep learning models that lack transparency and robustness \cite{zhou2022robustness}, {often ignoring the safety constraints or specifications}. Further, these approaches launch the recovery actions \textit{after} the attacks have already caused a noticeable deviation in the system states or resulted in hazards \cite{SAVIOR2020,choi_ccs_18}, which may be too late to prevent adverse events.




{
In addition to (i) strict timing constraints for anomaly detection and hazard prevention and (ii) application-specific safety requirements, major challenges impeding existing mitigation and recovery methods include (iii) the exponential growth of cyber-physical state spaces when generating corrective control action sequences and (iv) the lack of realistic closed-loop testbeds for safely simulating safety-critical fault scenarios and evaluating mitigation strategies without risking real-world harm.


In this paper, we focus on addressing these challenges in the context of APS, which is one of the most complex human-in-the-loop CPS. 
We propose \textit{KnowSafe},} a combined knowledge and data-driven approach for run-time prediction and mitigation of safety hazards resulting from critical attacks or accidental faults. \textit{KnowSafe} combines expert "Knowledge" on domain-specific "Safety constraints" with data from the closed-loop system operation to design a safety engine that can be integrated with a controller's interface to infer system context, predict impending hazards, and prevent the execution of unsafe control actions through generating preemptive and corrective actions. 


\begin{table}[t]
\centering
\caption{Recovery/Mitigation Work Comparison.}
\resizebox{0.9\columnwidth}{!}
{%
\begin{tabular}{|l|l|l|l|}
\hline
\textbf{Work} & \textbf{Attack Target} & \textbf{Miti. Method} & \textbf{Application} \\ \hline
\cite{dash2024specguard,pidpiper2021,liu2023learn} & Sensors & Data-Driven & \multirow{3}{*}{\begin{tabular}[c]{@{}l@{}}Drones/\\ Robotic\\ Vehicles\end{tabular}} \\ \cline{1-3}
\cite{Choi2020software,zhang2020real} & Sensors & Model-based &   \\ \cline{1-3}
\cite{ding2021mini} & Controller & Data-Driven &   \\ \hline
Ours & Controller & Combined & APS \\ \hline
\end{tabular}%
}
\vspace{-2em}
\label{tab:recoverywork}
\end{table}

The main contributions of the paper are as follows: 
\begin{itemize}[leftmargin=*]
    \item Proposing a novel approach for the integration of domain knowledge with recurrent neural networks (RNN) using customized loss functions to enforce application and context-specific safety constraints during model training. This approach enables more accurate and realistic prediction of system state sequences and hazards and the generation of corrective actions. 
    \item Developing a two-level regression RNN model to predict multi-variate system state sequences in both the far future for preemptive mitigation and the near future to ensure high accuracy. This approach relies on reachability analysis for hazard prediction, different from traditional ML anomaly detection methods that set a fixed threshold on the error between predicted and ground truth states (Section \ref{sec:predictnet}). 
    \item Proposing a knowledge-guided mitigation path planning algorithm, based on a variant of Rapidly-exploring Random Tree Star (RRT*) algorithm, that can find an optimal path for returning the system to the safe region of operation as soon as possible while satisfying application-specific constraints (Section \ref{sec:rrt}). 
    \item Developing a framework for formal specification of context-dependent mitigation requirements and their integration with ML models for the compliant and accurate generation of realistic response control actions (Section \ref{sec:ResponseAction}). 
    \item Evaluating the performance of the proposed hazard prediction and mitigation approach compared to state-of-the-art attack recovery methods using two actual clinical/patient datasets \cite{pso3-dataset,shahid2022large} and two closed-loop testbeds in APS with real-world control software, physical process (patient glucose) simulators, and a realistic adverse event simulator. 
    Experimental results show that \textit{KnowSafe} can accurately predict the occurrence of hazards with 28.1\% lower root mean square error (RMSE) than baselines while keeping the false positive rate low with no false negatives (Section \ref{sec:Evaluation:accuracy}). Further, it maintains the regular operation of simulated system despite faults or attacks and without introducing any new hazards, with a mitigation success rate of 92.8\%, which is at least 76\% higher than solely rule-based (50.9\%) and solely data-driven (52.7\%) methods (Section \ref{sec:Evaluation:pipeline}).

\end{itemize}

\section{Background}

CPS are closed-loop control systems designed by the tight integration of the inter-connected software and hardware components with sensors and actuators to monitor and control physical processes. A CPS controller adapts to the constantly changing and uncertain physical environment and the operator’s commands.




\begin{figure}
    \centering
        \includegraphics[width=0.6\columnwidth]{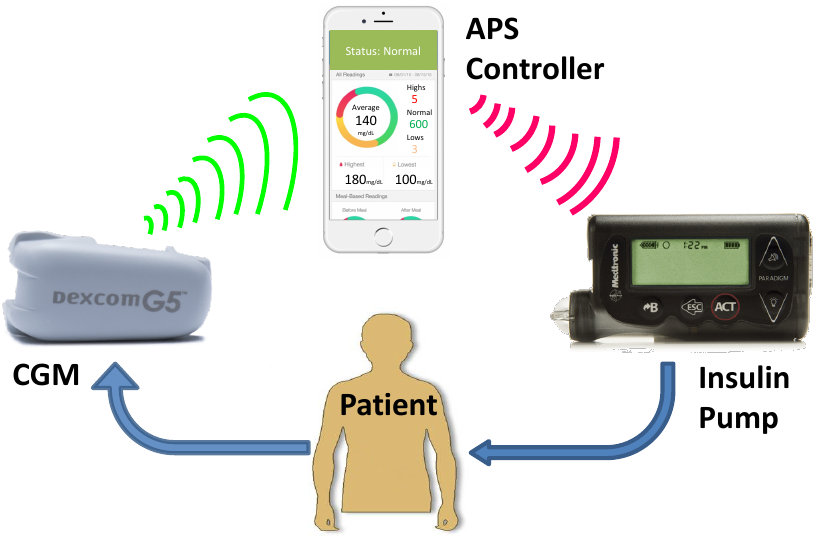}
        \vspace{-0.5em}
        \caption{Artificial Pancreas System (APS).} 
    \label{fig:apsstructure}
    \vspace{-1.5em}
\end{figure}


    

\subsection{Artificial Pancreas Systems}
Our main case study in this paper is Artificial Pancreas Systems (APS), an autonomous medical CPS used for the treatment of Type I diabetes. The structure of a typical APS is shown in Fig. \ref{fig:apsstructure}, including a continuous glucose monitor (CGM), an insulin pump, and a controller. At each control cycle, the APS controller estimates the current blood glucose (BG) level and insulin on board (IOB) based on measurements from the CGM sensor and decides on the dose of insulin to be injected through an insulin pump based on the patient's treatment targets, physical activities, and other parameters.  
IOB represents the remaining insulin in the patient's body, which will continue to digest blood glucose and should be considered for more accurate prediction of future BG trajectory as well as for the decision on insulin injections. 
Attacks or faults targeting any of the APS components, including the CGM, insulin pump, or the controller, can lead to underdose or overdose of insulin and risk of harm to patients \cite{li2014attacking, li2011hijacking, hackpump}. 
As shown in Fig. \ref{fig:apsstructure}, the modern APS controllers are implemented as mobile apps that communicate with the CGM and insulin pump wirelessly. Previous recalls and vulnerability reports have shown the possibility of the APS software and wireless link being compromised by attackers to threaten system security and patient safety (see Table \ref{tab:attackscenario}). 

\begin{table*}[t!]
\vspace{1em}
\centering
\caption{Simulated Attack Scenarios and Representative Real-world Vulnerabilities and Adverse Events Reported for APS.}
\label{tab:attackscenario}
\vspace{-0.5em}
\resizebox{\textwidth}{!}
{%
\begin{threeparttable}
\begin{tabular}{|l|l|l|l|l|}
\hline
\textbf{Attack Approach} & \textbf{Attack Target} & \textbf{Attack Examples} & \textbf{FDA Recalls$\textsuperscript{1}$}& \textbf{Reported Vulnerabilities $\textsuperscript{2}$} \\ \hline

Stop the insulin delivery & \multirow{2}{*}{Availability \cite{heartfield2018taxonomy}} & \cite{li2011hijacking, ramkissoon2017review,zhang2014trustworthiness} & Z-1890-2017, Z-2471-2018  & 
\multirow{4}{*}{\begin{tabular}[c]{@{}l@{}} Insecure wireless communication \cite{li2011hijacking,hackpump,hackpump2}, CVE\cite{CVE-2018-14781}\\ Lack of encryption or authorization \cite{availabilityattack, o2015cybersecurity}, CVE\cite{CVE-2019-10964}\\ 
Open source software \cite{asarani2021efficacy, openSourceOpenAPS} \\
Mobile/App based control \cite{eng2013promise,knorr2015security} \end{tabular}} \\  \cline{1-1}\cline{3-4}
Stop refreshing state variables & & \cite{glisson2015compromising,chacko2018security,giansanti2021cyber,sari2021health,flynn2020knock} & Z-0929-2020 &\\ \cline{1-4}  

Change insulin delivery amount & \multirow{2}{*}{Integrity \cite{mo2012integrity}} & \cite{li2011hijacking,hei2013pipac, marin2016feasibility,klonoff2015cybersecurity}  &Z-1583-2020, Z-1105-05  &\\ \cline{1-1}\cline{3-4}

Change controller input values &  & \cite{malasri2009securing,li2014attacking,kermani2013emerging,tamada2002keeping} & Z-0043-06, Z-1034-2015  & \\ \hline

\end{tabular}%

\begin{tablenotes}
     \item[1] Recall IDs are assigned by the U.S. FDA and can be searched for on \url{https://www.accessdata.fda.gov/scripts/cdrh/cfdocs/cfres/res.cfm}.
     \item[2] Vulnerabilities are identified from the related works and by searching the national vulnerability database using relevant keywords (e.g., insulin pump) \cite{CVE}.
    \end{tablenotes}
    
    \end{threeparttable}
}
\vspace{-1em}
\end{table*}

\subsection{Control System Model}
At each control cycle $t$, a CPS controller estimates the system's current state $x_t=(x^1_t,x^2_t,...,x^p_t)$ based on sensor measurements, decides a control action $u_t$ to change the state of the physical system based on the current state and control targets, and sends $u_t$ as a control command to the actuators. Upon execution of the $u_t$ by the actuators, the physical system will transition to a new state $x_{t+1}$. 

We consider three typical regions of operation in the state space of a CPS controller.
The unsafe/hazardous region ${\mathcal{X}}_{h}$ is defined as the set of states that lead to adverse events (e.g., the set of states corresponding to hyperglycemia {(BG>180mg/dL)} and hypoglycemia events {(BG<70mg/dL) in APS}). The safe/target region ${\mathcal{X}}_{*}$ is defined based on the goals and guidelines of the specific application (e.g., keeping BG in the target range of 120-150 mg/dL). The set of states not included by either of these regions is referred to as the possibly hazardous region ${\mathcal{X}}_{*<h}$. 

\begin{figure}[b]
    \centering
    \vspace{-1.5em}
    \includegraphics[width=.65\columnwidth]{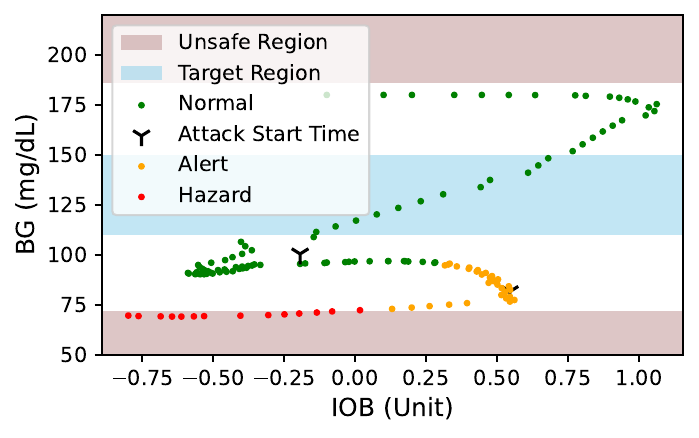}
    \vspace{-1em}
    \caption{Example Regions of Operation and an APS System State Trajectory under Attack.} 
    \label{fig:regionexample}
\end{figure}

For modeling the multi-dimensional context in CPS (including physical processes and environment that affect the cyber processes) and the human-interpretable specification of their interactions in both temporal and spatial domains, 
we define $\mu(x_t) = (\mu_{1}(x_t),\dots, \mu_{m}(x_t)) \in \mathbb{R}^m$, where $\mu_{i}(x_t)$ are transformations of $x_t$, (e.g., derivative, polynomial combinations, or other functions) modeling more complex combinations of state variables and their rates of change.
The set of all possible values of $\mu(x_t)$ is denoted by $\mathcal{M}$. We describe the \textit{system context} $\rho(\mu(x_t))$ as subsets of $\mathcal{M}$, defined by ranges of variables in $\mu(x_t)$, that can be mapped to the regions $\{{\mathcal{X}}_{*},{\mathcal{X}}_{*<h}, {\mathcal{X}}_{h}\}$. 
To identify the nature of a control action $u_t$ within a context $\rho(\mu(x_t))$, we need to determine the possibility that by issuing $u_t$ the system eventually transitions into a new context $\rho(\mu(x_{t'}))$ within the hazardous region ${\mathcal{X}}_{h}$. Fig. \ref{fig:regionexample} shows example regions of operation for APS and a state trajectory where an attack on controller state variables causes the system's transition into the unsafe region.

\subsection{Threat Model}
\label{sec:threatmodel}

We focus on accidental faults or malicious attacks targeting the CPS \textit{controller} that, if activated under certain triggering conditions, can manifest as errors in the inputs, outputs, or internal state variables of the control software and cause unsafe control actions and hazardous system states, which, together with a worst-case set of environmental conditions, could lead to adverse events \cite{leveson2013stpa}, such as those reported in the literature or public datasets, e.g., severe complications from hypoglycemia or hypoglycemia~\cite{brown2019six,pedersen2017reporting}. A comparison between our work and existing mitigation solutions in CPS is shown in Table \ref{tab:recoverywork}.
We assume such errors are transient and only occur once for a particular duration.
We aim to mitigate these faults/attacks by predicting hazardous system states and correcting unsafe control actions. 
%


The attackers can change the controller state through external or insider threats. 
For example, they can modify the values of the controller state variables by exploiting the known vulnerabilities in the communication channels of devices \cite{alemzadeh2016targeted,ramkissoon2017review, CVE-2019-10964}, underlying operating system (e.g., malicious wrappers for system libraries) \cite{alemzadeh2016targeted}, or control-semantic bugs (e.g., input validation bugs) \cite{ding2021mini}.
The increasing popularity of open-source \cite{opensourcechallenge} and mobile app-based  \cite{eng2013promise,dehling2015exploring} controllers and wireless/radio communications \cite{FDApumprecall} further raises the possibility of such attacks. 
The attackers can also obtain unauthorized remote access to the control software by exploiting weaknesses such as stolen credentials \cite{Stuxnet}, vulnerable services \cite{Zetterhackhospital}, and unencrypted communication \cite{zhou2022Strategic}. 
Even for a  control system that does not connect to a network, the attacker can still exploit USB ports or Bluetooth to get one-time access and deploy malware. Table \ref{tab:attackscenario} presents the simulated attack scenarios in this paper and representative real-world vulnerabilities and recalls reported to the U.S. Food and Drug Administration (FDA) for APS.


We assume that the sensor measurements received by the safety engine and mitigation modules are not compromised either because the attacks directly target the controller software without affecting sensor data or sensor attacks are detected and prevented from propagation to the safety engine using one of the well-studied sensor anomaly detection solutions~\cite{Choi2020software,SAVIOR2020,choi_ccs_18,attack_pcs,luo_deep_2021}. 
In Section \ref{sec:sensorattack}, we conduct experiments on how the performance of such sensor anomaly detection methods can affect the mitigation performance.
We assume that the safety engine is made tamper-proof (e.g., using protective memories or hardware isolation \cite{TrustZone}) {add placed in the final computational layer (e.g., inside the pump), which is closer to the actuator and harder to attack}. The safety engine is integrated with the CPS controller as a wrapper with only access to the input and output data without any changes to the original system architecture. 
{Further study on mitigating attacks that can evade existing sensor anomaly detection methods or compromise the safety engine is beyond this paper's scope.}


\subsection{Design Challenges}
\label{sec:DesignChallenge}
As mentioned in the introduction, the existing works on enhancing APS security mainly focus on encryption \cite{weng2023ensuring}, intrusion detection \cite{aliabadi2021artinali}, safety specification \cite{astillo2021smdaps}, and model-based \cite{vega2009increasing} or data-driven \cite{dsn2021zhou} run-time monitoring with less attention paid to hazard mitigation.
The only work we found that attempts to avoid hazards in APS includes a fail-safe approach that shuts down the pump or stops insulin delivery under attack \cite{zhang2011generic,paul2011review}. However, this approach fails to maintain the regular operation of APS. 

The major challenges impeding the effectiveness of existing mitigation and recovery methods include (C1) strict timing constraints for taking corrective actions and preventing adverse events, (C2) cyber-physical state explosion in reachability analysis and the generation of corrective control action sequences, and (C3) lack of realistic closed-loop testbeds for simulation of safety-critical fault/attack scenarios and testing the mitigation strategies without the risk of adverse consequences.

\begin{figure}[t]
    \centering
    \includegraphics[width=0.8\columnwidth]{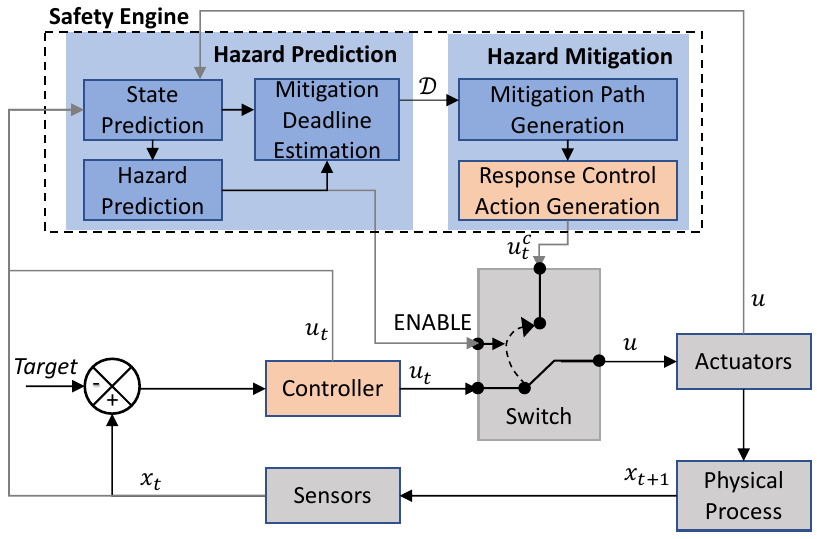}
    \vspace{-0.5em}
    \caption{Combined Knowledge and Data Driven Safety Engine for Hazard Prediction and Mitigation.}
    \label{fig:framework}
    \vspace{-1.5em}
\end{figure}

\section{Combined Knowledge and Data Driven Hazard Mitigation Approach}
\label{sec:design}
To overcome the aforementioned challenges, this paper proposes a combined knowledge and data driven approach for designing a safety engine that can predict and mitigate potential hazards at run-time. 
Fig. \ref{fig:framework} shows the overall design of the safety engine, consisting of a \textit{hazard prediction} module (Section \ref{sec:predictnet}) and a hazard mitigation module, with \textit{mitigation path planning} (Section \ref{sec:rrt}) and \textit{corrective control action generation} (Section \ref{sec:ResponseAction}). 

At each control cycle, the hazard prediction module takes as input a sequence of system states or their transformations as inferred from sensor data and precisely estimates their future trend and trajectory for hazard prediction (C1). The time between the current time and when the inferred hazard will happen is defined as the \textit{mitigation deadline} or the budget to launch corrective or response control actions and mitigate hazards. If any potential hazards are inferred, the hazard mitigation module generates an optimal path and a corresponding control action sequence to keep the system away from the unsafe region and return it to the target region under the constraint of mitigation deadline and other domain-specific safety requirements (C2). 


\subsection{Reachability Analysis and Hazard Prediction}
\label{sec:predictnet}
Anomaly detection is a critical step in hazard mitigation and recovery. Previous works on anomaly detection and recovery mainly focus on comparing the predicted states with actual states based on sensor data and waiting until a large deviation over a detection window to raise an alarm \cite{Choi2020software,SAVIOR2020}. However, relying on predefined thresholds for detecting deviations may lead to delayed detection and high false positive rates, resulting in mitigation failures. In this work, we propose a hazard \textit{prediction} method that estimates the possible sequence of future system states and whether they will fall within any unsafe regions of operation. 

\textbf{Problem Statement:}
We specifically define the following binary classification problem: 
\begin{equation}    
y_t=
\begin{cases}
1, & \text{ if } \exists t'\in [t,t+N]: \{ \hat{x}_{t'}\in g(X_{t},U_{t})\} \subset  {\mathcal{X}}_{h} \\
0, & \mathrm{otherwise} 
\end{cases}
\end{equation}
\noindent where $g(\cdot)$ is a prediction model that given an input state sequence $X_t=\{x_{t-k},...,x_t\}$ and the control action sequence $U_t=\{u_{t-k},...,u_t\}$, predicts the system state trajectory $\hat{X}_{t+N}=\{\hat{x}_{t+1},...,\hat{x}_{t+N}\}$ within a prediction window of $N$ control cycles. A hazard is predicted to occur if any predicted state $\hat{x}_{t+i}$ is located within the unsafe region $\mathcal{X}_h$. The mitigation deadline, $\mathcal{D}$, is further estimated by calculating the difference between current time $t$ and the minimum $t'$ within the prediction window such that the system state $\hat{x}_{t'}$ is unsafe.

\vspace{-1em}
\begin{equation}
\mathcal{D} = min\{t' \in [t,t+N]\}-t : \hat{x}_{t'} \in \mathcal{X}_h 
\end{equation}

\textbf{Domain Knowledge Specification:}
To ensure that the predicted system state sequences are accurate and realistic, we need to check a set of application-specific constraints, e.g., whether the change from the start state to the predicted states is within a reasonable range. 
These properties are often described in human-interpretable format over a transformed state space representing the derivatives, polynomial combinations, or other functions of state variables (e.g., the first derivative of blood glucose to represent the rate of change). So we project the current state $x_{t+i}$ and its prediction in the next step $\hat{x}_{t+i+1}$ to the transformed state space of $\mu(x)=(\mu_1(x),\mu_2(x),...,\mu_m(x))$. 
Then, we define the $\delta$ reachable state of $\mu(x_t)$ as a set or region $\mathcal{S}(\mu(x_t),\delta)$, such that the difference between any states within this region $\mathcal{S}$ and the state $\mu(x_t)$ is constrained by the parameter set $\delta$ as shown in Eq. \ref{eq:reachablestate}:
\begin{align}
\label{eq:reachablestate}
    \mathcal{S}(\mu(x_t),\delta):=\{\forall s_t \in \mathcal{S}, |s_t - \mu(x_t)| \leq \delta \}
\end{align}

\noindent where $\delta=(\delta_1,\delta_2,...,\delta_m)$ is a set of parameters for each variable in $\mu(x)$ and could be derived based on domain-specific guidelines, design requirements, practical experience, or statistical analysis of the past data. 

\begin{figure}[t]
    \centering
    \includegraphics[width=\columnwidth]{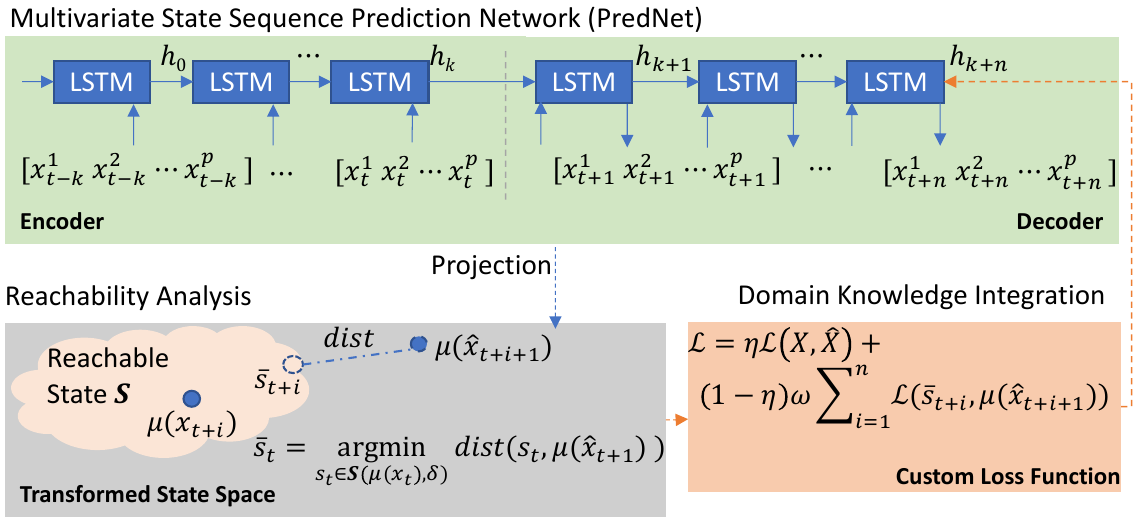}
    \caption{Multivariate Regression Neural Net for Sequential States Prediction (PredNet) with Domain Knowledge Integration.}
    \label{fig:seq2seq}
    \vspace{-2em}
\end{figure}

\textbf{Data and Knowledge Integration:}
In this paper, we develop a novel multivariate regression neural network model (referred to as \textit{PredNet}) to predict the sequence of system states using either the data collected from closed-loop CPS simulations or real system operation (e.g., from clinical trials as described in Section \ref{sec:Evaluation:accuracy}). As shown in Fig. \ref{fig:seq2seq}, our prediction network is based on an encoder-decoder model that predicts the future system states $x_{[t+1:t+N]}$ for $N=n$ steps based on a sequence of $k+1$ past system states.

%

The domain-specific properties are integrated with \textit{PredNet} by adding a customized loss function as the regularization term to the original loss function $\mathcal{L}(X,\hat{X})$ (e.g., mean squared error), as shown below:

\vspace{-1em}
\begin{equation}
\label{eq:loss_reachablestate}
   \mathcal{L}=\eta\mathcal{L}(X,\hat{X}) + (1-\eta)\omega\sum_{i=1}^{n}\mathcal{L}(\bar{s}_{t+i},\mu(\hat{x}_{t+i+1}))
\end{equation}

\begin{equation}
    \bar{s}_t = \underset{s_t \in \mathcal{S}(\mu(x_t),\delta)}{\mathrm{argmin}} dist(s_t,\mu(\hat{x}_{t+1}))
\end{equation}
\noindent where, $\bar{s}_{t+i}$ is the state within the $\delta$ reachable state of $\mu({x}_{t+i})$ that has the minimum distance (e.g., Euclidean distance) to the predicted state transformation $\mu(\hat{x}_{t+i+1})$, and $\omega$ is a weight parameter that takes a zero value when $\mu(\hat{x}_{t+i+1})$ is within the reachable state $S$ and a unit value otherwise. $\eta \in [0,1]$ represents the weight of original loss contributing to the total loss. In this paper, we select the value of $\eta$ by manual hyper-parameter tuning. The custom loss function enforces realistic trajectories by penalizing \textit{PredNet} during training whenever a predicted state is outside the reachable set of the current state. 

We also design a two-level hazard prediction pipeline consisting of (i) a long-term prediction model (\textit{PredNet-l}) that can predict a far-future state sequence with the benefit of ensuring enough time to foresee and mitigate potential hazards, and (ii) a short-term prediction model (\textit{PredNet-s}) that has higher accuracy than the long-term model, ensuring we do not miss any hazards.

\subsection{Mitigation Path Generation}
\label{sec:rrt}
The next major step in hazard mitigation is the selection of response actions that the system should take to complete its mission (i.e., end up or stay within the target region for the controller) safely. 
A simple algorithm to execute the proposed mitigation might involve control action compensation based on fixed rules or medical guidelines (e.g., increase insulin injection to prevent hyperglycemia in APS).
However, such a generic mitigation mechanism does not consider the current context of the system or its interactions with the environment and might lead to over-/under-mitigation or potential harm to system users~\cite{dsn2021zhou}.
It is challenging to design a mitigation mechanism with a high recovery rate that introduces as few new hazards as possible.

Our goal is to learn a sequence of control actions that can mitigate hazards and keep the system away from unsafe regions.
Simply, we could feed all the possible actions to the prediction model introduced above and select the optimal one that results in a future system state sequence closest to the safe region. However, this approach would be time- and resource-consuming as we must traverse the whole action space and may not converge to an optimal solution within the required mitigation deadline. 
Therefore, instead of generating the recovery control actions directly, we first find the desired sequence of recovery states (using a fast path planning algorithm) and then use a control algorithm (ML-based or model-based) to generate a corresponding sequence of control actions. 

\textbf{Problem Statement:}
We model the mitigation path generation as an optimization problem with the goal of returning the system to the safe region as fast as possible and before the occurrence of hazards while avoiding the unsafe region. Specifically, we minimize the time to mitigate the hazard {(refereed to as $TTMH$)} under the constraints shown in the following equations:

\vspace{-1.5em}
\begin{align}
\label{eq:opt}
    \text{min} &\{TTMH\};\ \text{s.t.}\ \\
    %
    & {\Delta x_i}/{T} < {\alpha}_i^1; \forall i \in [1,p] \label{eq:limit1}\\
    %
    & {\Delta(\Delta x_i)}/{T^2} < {\alpha}_i^2 ; \forall i \in [1,p] \label{eq:limit2}\\
    & \{x\} \cap \mathcal{X}_h = \varnothing \\ 
    & PathLength < \mathcal{D} \label{eq:limit3}
\vspace{-2em}
\end{align}

\noindent where, $x$ is the system state, $T$ is the duration of a control cycle, and $\alpha_i^j$ imposes application-specific constraints {on the first and second derivatives, such as limiting the rate and acceleration/smoothness of blood glucose changes}. $\mathcal{D}$ is the maximum mitigation budget to mitigate hazards and the maximum length of mitigation path ($PathLength$), estimated using the hazard prediction module (see Fig. \ref{fig:framework}). {To ensure safety, the system state must maintain a separation from the unsafe region, represented by ensuring the empty set \( \varnothing \) remains between them (Eq. 9).}

\textbf{Domain Knowledge Specification:} 
The constraints ${\alpha^j_i}$ ensure satisfying the safety proprieties and keeping the smoothness between state transitions and are specified based on domain knowledge or practical experience. For example, the change of BG in APS should not exceed [-5,3] $mg/dL$ per 5 minutes \cite{dsn2021zhou}.

\textbf{Data and Knowledge Integration:}
We develop a mitigation path planning algorithm with the knowledge of safety constraints (referred to as \textit{SC-RRT*}) based on a variant of a path planning algorithm called Rapidly-exploring Random Tree Star (RRT*) \cite{karaman2011sampling}, which can easily handle problems with obstacles and differential constraints with a wide search range, fast search speed, and high computational efficiency and has been widely used in robotic motion planning. 
\textit{SC-RRT*} is a lightweight algorithm with more transparency than neural networks and more suitable for implementation on devices with limited memory and computation resources.
Algorithm \ref{alg:rrt*} shows the overall mitigation path generation process.
At each exploration cycle, the algorithm randomly samples a point in the state space (line 17) and configures a new vertex in the random tree $G$ in the direction from the nearest vertex not exceeding the incremental distance $\Delta x$ (lines 18-19). 
We encode the domain knowledge as safety constraints to filter out the invalid vertices in the random sampling step.
Specifically, a random-sampled vertex is only added to the tree if its connection to the nearest node satisfies the safety requirements defined in Eq. \ref{eq:limit1}-\ref{eq:limit3}. The edges between the new node and its neighbor nodes within a radius $\gamma$ are further updated if any lower-cost connects are found (lines 22-27). A mitigation path is found if the destination locates inside the target region $\mathcal{X}_*$ (lines 28-29). Finally, an optimal destination vertex and mitigation path are selected (line 31) with the shortest path from the current state to the target state while optimizing other application-specific requirements (e.g., maximizing the smoothness or minimum distance to $\mathcal{X}_h$). 

\begin{algorithm}[t]
\footnotesize
 \DontPrintSemicolon
    \caption{Mitigation Path Planing Algorithm}
    \label{alg:rrt*}

    \KwIn {Initial config $x_{init}$, Number of vertices $K$, Incremental distance $\Delta$x, Near Radius $\gamma$, Recovery Budget $D$}

    \KwOut {RRT graph G, Optimal Goal Vertex $x_{optimal}$ , Minimum recovery path $path_{min}$}
    G.init($x_{init}$)  {\Comment{Init Optimal RRT}}\\
    $X_{dest}$ $\gets$ $\emptyset$ \Comment{Vertices in Target Region} \\

    \SetKwFunction{FUpdateCost}{UpdateCost}
    \SetKwProg{Fn}{Function}{:}{}
    \Fn{\FUpdateCost{$X_{near}$, $x_{new}$}}{
        \For{$x_{near} \in X_{near}$} 
            {
                \uIf{ValidConnection($x_{near},x_{new}$)}
                {
                    \uIf{Cost($x_{near}$)+Dist($x_{nearest},x_{new}$) < $cost_{min}$}{
                    $x_{min} \gets x_{near},cost_{min}$ $\gets$ \\ Cost$(x_{near})$+Dist($x_{nearest},x_{new}$) \\
                }
                }
            }
    }
    \KwRet

    \SetKwFunction{FRewriteTree}{RewriteTree}
    \Fn{\FRewriteTree{$X_{near}$, $x_{new}$}}{
        \For{$x_{near} \in X_{near}$}{
        
            G.update($G, x_{near}, x_{new}$) 
            \Comment{update shorter path with $x_{new}$}
        }
    }
    \KwRet

    \For{k = 1 to K}{
         $x_{rand}$ $\gets$ RandomSampling() \\
         $x_{nearest}$ $\gets$ FindNearestNode(G,$x_{rand}$)\\
         $x_{new}$  $\gets$ NewConf($x_{nearest}$, $x_{rand}$, $\Delta$ x)\\ 
         
         \uIf{ValidConnection($x_{nearest}$,$x_{new}$)}
         {
            $G.add\_node(x_{new})$\\
            $X_{near} \gets$ FindNearNodes(G,$x_{new}$,$\gamma$)\\
            $x_{min} \gets x_{nearest}$ \\
            $cost_{min} \gets$ Cost($x_{nearest}$)+Dist($x_{nearest},x_{new}$)\\
            
            UpdateCost($X_{near}$, $x_{new}$) \\
            
            G.add\_edge($x_{min},x_{new}$) {\Comment{Shortest path to $x_{new}$}}\\
            RewriteTree($X_{near}$, $x_{new}$) 

         }
        
        \uIf{$x_{new}$ $\in$ Goal}
        {
            $X_{dest} \gets X_{dest} \cup x_{new}$
        }
    }
    $x_{optimal},path_{min} \gets$ FindOptimalNode($G, \mathcal{X}_{*}$) 
    
    \Return{G, $x_{optimal}$, $path_{min}$}
\end{algorithm}
\vspace{-1em}

\subsection{Response Control Action Generation}
\label{sec:ResponseAction}




Once the target state trajectory is generated, it can be fed to a control algorithm (either ML-based or model-based) to generate the final sequence of corrective/recovery control actions that realize the desired state trajectory.  

\textbf{Problem Statement:}
We model the task of generating a sequence of control actions as a context-specific multivariate sequence-to-sequence regression problem, as shown below: 
\begin{align}
\label{eq:ml-model0}
&y_{t} =f(\Bar{X}_{t+n})= \hat{U}_{t+n-1}, \text{s.t.}\\
& \forall t'\in [t,t+n-1]: x_{t'}\overset{u_{t'}}{\rightarrow}x_{t'+1} \nonumber
\end{align}

Given the expected system state sequence ${\Bar{X}}_{t+n}=\{\Bar{x}_{t+1},...,\Bar{x}_{t+n}\}$, the controller outputs a control action sequence $\hat{U}_{t+n-1}$ = $\{\hat{u}_{t}$, $...$, $\hat{u}_{t+n-1}\}$ that if executed sequentially by the actuators should transition the system to the expected states.  

\textbf{Domain Knowledge Specification:}
\label{sec:HMS}
%
The goal of mitigation is that given the current system context and prediction of a potential hazard, a new corrective control action $u^c_t$ from a set of control actions $\boldsymbol{u}^\rho$ is sent to the actuators to force the system transition into the target or safe region. The context-specific corrective control actions could be specified based on domain knowledge. 
Here, we use a formal framework based on Signal Temporal Logic (STL) for the human-interpretable description of the mitigation actions. 
STL is a formal specification and machine-checkable language often used for rigorous specification and run-time verification of requirements with time constraints in CPS \cite{bartocci2018monitoring}. 
For the system context $(\rho(\mu(x_{t}))$, we describe the Hazard Mitigation Specification (HMS) $(\rho(\mu(x_{t})), u^c_t)\mapsto {\mathcal{X}}_{*}$ as follows: 
\begin{equation}
\Scale[0.92]{
    G_{[t_{0},t_{e}]}( (F_{[0,t_{s}]}(u^c_t))\mathcal{S}(\varphi_{1}(\mu_{1}(x_{t})) \land \ldots \land  \varphi_{m}(\mu_{m}(x_{t}))))
    \label{eq:HMS}
}
\end{equation}

\noindent where $F$ is the eventually operator requiring that $u^c_t \in \boldsymbol{u}^{\rho}$ should be taken within period $t_{s}$ since (denoted by the $S$ operator) the system enters context $\rho(\mu(x_{t}))$ measured by $(\varphi_{1}(\mu_{1}(x_{t})) \land \ldots \land  \varphi_{m}(\mu_{m}(x_{t})))$. 
Each $\varphi_{i}(\mu_{i}(x_{t}))$ is an atomic predicate that represents an inequality on $\mu_{i}(x_{t})$ in the form of 
$ \mu_{i}(x_{t}) \{<, \leq , >, \geq\} {\beta}_{i}$ or its combinations, where the inequality thresholds ${\beta}_{i}$ define the boundary of the subset in that dimension $\rho(\mu_{i}(x_{t}))$.
This should hold globally (denoted by the $G$ operator) between the start time $t_{0}$ and the end time $t_{e}$ of the system operation.
The time parameter $t_s$ specifies the requirement for the latest possible time a mitigation action should be initiated after a potential unsafe control action is detected to prevent hazards. 
The estimated time between the activation of a fault/attack on the system and the occurrence of a hazard can provide an upper bound for specifying this time requirement.

The formal safety specifications are usually generated by safety analysts in cooperation with domain experts using hazard analysis approaches \cite{dsn2021zhou,zhou2023hybrid}.
In this paper, we adopt a control-theoretic hazard analysis method (called STPA \cite{leveson2013stpa}) to generate HMS by following these steps:
\begin{enumerate}
    \item Define the hazards and adverse events of interest.

    \item Identify the observable set of variables $x_{t}$ of interest related to the hazards and decide on the possible transformations $\mu(x_{t})$ and the sets $\rho(\mu(x_{t}))$ as completely as possible. 
    \item List all the combinations of $\rho(\mu(x_{t}))$ and control action $u_{t}$ and identify the combinations that might result in transitions to a hazardous region ${\mathcal{X}}_{h}$. The unknown boundaries $\beta_i$ can be learned from a population dataset using off-the-shelf optimizers (e.g., Newton's method) \cite{dsn2021zhou,zhou2023hybrid}.
    \item For each unsafe context in step 3, find all control actions $u^c_t \in U$ such that $(\rho(\mu(x_{t})),u^c_t) \mapsto {\mathcal{X}}_{*}$ and add these tuples $(\rho(\mu(x_{t})),u^c_t)$  to HMS for that context. 
    \item Formalize the generated HMS into STL format.
\end{enumerate}
An example of generated STL hazard mitigation specification for APS is shown in Table \ref{table:stltable}, where a set of STL rules are specified to mitigate potential hazards that may result in hyperglycemia or hypoglycemia events. 
For example, the first rule states that since the system context that BG is larger than the target and keeps increasing and IOB is less than a threshold $\beta_1$ and keeps decreasing, an \textit{increase\_insulin} action $u_2$ should be issued by the controller within time $t_s$ to avoid hyperglycemia event. This safety property should be held true during the whole operation period.

\begin{table}[t]
\centering
\vspace{1em} 
\caption{STL Hazard Mitigation Specifications for APS.}
\vspace{-1em}
\label{table:stltable}
\resizebox{\columnwidth}{!}{%
\begin{threeparttable}

\begin{tabular}{|c|ll|} \hline
\textbf{Rule} & \textbf{Mitigation Action}&\textbf{STL Description of Safety Context} \\ \hline

1 & ${G_{[t_0,t_e]}((F_{[0,t_{s}]}(u_2))}$&$\mathcal{S}((BG>BGT\wedge BG'>0)\wedge (IOB'<0\wedge IOB<\beta_{1})))$      \\ \hline 
2 & ${G_{[t_0,t_e]}((F_{[0,t_{s}]}(u_2||u_4))}$&$\mathcal{S}((BG>BGT\wedge BG'>0)\wedge (IOB'=0\wedge IOB<\beta_{2})))$ \\ \hline 
3 & ${G_{[t_0,t_e]}((F_{[0,t_{s}]}(u_2||u_4))}$&$\mathcal{S}((BG>BGT\wedge BG'<0)\wedge (IOB'>0\wedge IOB<\beta_{3})))$ \\ \hline 
4 & ${G_{[t_0,t_e]}((F_{[0,t_{s}]}(u_2||u_4))}$&$\mathcal{S}((BG>BGT\wedge BG'<0)\wedge (IOB'<0\wedge IOB<\beta_{4})))$ \\ \hline 
5 & ${G_{[t_0,t_e]}((F_{[0,t_{s}]}(u_2||u_4))}$&$\mathcal{S}((BG>BGT\wedge BG'<0)\wedge (IOB'=0\wedge IOB<\beta_{5})))$ \\ \hline 
6 & ${G_{[t_0,t_e]}((F_{[0,t_{s}]}(u_3))}$&$\mathcal{S}((BG<BGT\wedge BG'<0)\wedge (IOB'>0\wedge IOB>\beta_{6})))$      \\ \hline 
7 & ${G_{[t_0,t_e]}((F_{[0,t_{s}]}(u_1||u_3))}$&$\mathcal{S}((BG<BGT\wedge BG'<0)\wedge (IOB'<0\wedge IOB>\beta_{7})))$      \\ \hline 
8 & ${G_{[t_0,t_e]}((F_{[0,t_{s}]}(u_1||u_3))}$&$\mathcal{S}((BG<BGT\wedge BG'<0)\wedge (IOB'=0\wedge IOB>\beta_{8})))$      \\ \hline 
9 & ${G_{[t_0,t_e]}((F_{[0,t_{s}]}(\neg u_3))}$&$\mathcal{S}((BG>BGT\wedge IOB<\beta_{9})))$                            \\ \hline 
10 &${G_{[t_0,t_e]}((F_{[0,t_{s}]}(u_3))}$&$\mathcal{S}((BG<\beta_{10})))$                                              \\ \hline

\end{tabular}

\begin{tablenotes}\footnotesize
\item[*] BGT: BG target value, IOB: Insulin on board;
$BG' =dBG/dt$, $IOB'=dIOB/dt$; 
\item[*] $u_{1,2,3,4}:$decrease\_insulin, increase\_insulin, stop\_insulin, keep\_insulin;

\end{tablenotes}

\end{threeparttable}
}
\vspace{-1em}
\end{table}
It should be noted that there is a semantic gap between the high-level human-interpretable state variables $\mu(x)$ in the HMS (e.g., rate of BG change) and the sensor measurements (e.g., BG values from CGM) as well as the high-level control actions $u_t^c$ (e.g., \textit{increase insulin}) and the low-level control commands sent to the actuator (e.g., the amount of insulin dose). To close this gap during actual implementation, an additional step is needed to infer the values of state variables in the transformed state space based on the sensor measurements (e.g., by taking the derivative of BG values to calculate the rate of BG change) and the high-level control commands based on the ML predictions, using the transformation function $h(\cdot)$ (see Eq. \ref{eq:loss_custom}). These transformations enable matching of the estimated states and control actions with the logic formulas specified in HMS to check the satisfaction of any of the safety properties (e.g., Rules 1-5 in Table \ref{table:stltable} will be checked for control action $u_{2}$).


\textbf{Data and Knowledge Integration:}
In this paper, we integrate the domain-specific knowledge on context-dependent mitigation actions with state variable and control output data traces collected from closed-loop CPS simulation or real operation to train an RNN for control action generation (referred to as \textit{ActNet}). 
We encode the logic formulas generated for HMS as a custom loss function that penalizes the RNN model during the training process if the ML prediction does not match the specified mitigation action properties:
\vspace{-0.25em}
\begin{equation}
    \vspace{-0.25em}
    \label{eq:loss_custom}
    \mathcal{L} = \eta\mathcal{L}(U,\hat{U}) + (1-\eta)k(2\sigma(\sum_{h(y_{t}) \notin u_t^c} \omega_i) -1)
\end{equation}


\noindent where, $h(\cdot)$ is a transformation function (e.g., derivation) that transfers the predicted control command $y_t$ to a discrete high-level control action as described above, and 
$w$ is a weight parameter that defines the impact of each logic rule on the training process. $\sigma(\cdot)$ is a sigmoid function that maps the degree of satisfaction of the ML outputs with the STL formulas to the range of [0,1], and
$k$ is a scaling factor that scales the sigmoid function to a similar range with the original loss $\mathcal{L}(U,\hat{U})$ (e.g., mean squared error).

%

\subsection{Safety Engine Implementation}
\label{sec:Implementation}
\begin{algorithm}[t]
\footnotesize
\caption{Hazard Prediction and Mitigation}
\label{alg:mitigation}

\SetKwFunction{FMain}{Mitigation\_Process}
    \SetKwProg{Fn}{Function}{:}{}
    \Fn{\FMain{}}{
    \eIf{$MitEnable$ }{ 
        \eIf{$x_t \in \mathcal{X}_*$ {\bf or} $kMit \geq \mathcal{D}$}{
            $MitEnable \gets False$ \Comment{Stop } \\
            $kMit \gets 0$ \\
            \uIf{$x_t \notin \mathcal{X}_*$}{
            RaiseAlert("Mit. Failure!")\\
            }
        }
        {
            \uIf{$kMit>0$ {\bf and} $err({\Bar{X}}_{[kMit-1]},x_t)>\theta$}{
                 
                ${\Bar{X}} \gets SC\_RRT^*(x_t,\mathcal{D}-kMit,QuickMit) $ \\
                $U^c \gets ActNet({\Bar{X}})$  \Comment{Update}\\
                $kMit \gets 0$; 
                $\mathcal{D} \gets \mathcal{D}-kMit $ \\
                RaiseAlert("Mit. Plan Updated!")\\
            }

            $u \gets GetAction(U^c,kMit)$ \Comment{Replace}  \\  
    
            $kMit \gets kMit+1$ \\
        }
    }
    {
        $u\gets u_t$ \\
    }
    Actuate($u$)
    }
\KwRet

$x \gets$ \text{System State}\\
$u \gets$ \text{Control Action to the Actuator}\\ 
$Hazard \gets False$ \Comment{Hazard Prediction Flag}

\For{\text{each control cycle} t}{
    $X_t \gets UpdateState(x_t)$\\
    $U_t \gets UpdateAction(u_t)$ \Comment{Original Control Action}\\ 

    \If{MitEnable==False}{
    $kMit \gets 0$ \\
    $\hat{x}_{[t+1:s]} \gets PredNet\textit{-}s(X_t,U_t,s)$ \Comment{Short-term}\\ 
    \uIf{$\{\hat{x}_{[t+1:s]}\} \cap \mathcal{X}_h$}{
        $MitEnable \gets True$ \\
        $\mathcal{D} \gets Deadline(\hat{x}_{[t+1:s]})$ \\ 
        $QuickMit \gets True$ \\
    }
    \Else{
        $\hat{x}_{[t+1:l]} \gets PredNet\textit{-}l(X_t,U_t,l)$ \Comment{Long-term}\\ 
        
        \uIf{$\{(\hat{x}_{[t+1:l]}\} \cap \mathcal{X}_h$}{
            $MitEnable \gets True$ \\
            $\mathcal{D} \gets Deadline(\hat{x}_{[t+1:l]}) $ \\
            
            $QuickMit \gets False$ \\
        }
    }
    \uIf{MitEnable }{ 
        ${\Bar{X}} \gets SC\_RRT^*(x_t,\mathcal{D},QuickMit)$  \\
        \uIf{$\Bar{X}==\emptyset$}{
            RaiseAlert("Mit. Path Not Found!")\\
            \textbf{break}
        }
        $U^c \gets ActNet({\Bar{X}})$ \Comment{Corrective Action Seq.}\\
        {RaiseAlert("Hazard Detected. Please Perform Mit. Immediately!")}\\
        {RaiseAlert("Suggested Mit. Plan Generated!")}\\
    }

    }

            

    

        Mitigation\_Process()
      
}
\end{algorithm}

The overall hazard prediction and mitigation procedure is shown in Algorithm \ref{alg:mitigation}. 
Our two-level hazard prediction framework consists of (i) a long-term prediction model (\textit{PredNet-l}) that can predict a far-future state sequence with the benefit of ensuring enough time to foresee and mitigate potential hazards (line 36), and (ii) a short-term prediction model (\textit{PredNet-s}) with higher accuracy than the long-term model, ensuring we do not miss any hazards (line 30). 
The control system will raise an alert and switch to the \textit{Mitigation} mode when any predicted system state sequence overlaps the unsafe region $\mathcal{X}_h$ (lines 31-32 and lines 37-38). To reduce computation time (see Section \ref{sec:utilization}), \textit{PredNet-l} is executed only when \textit{PredNet-s} does not infer a hazard in a near future.

An optimal mitigation path is generated based on the remaining mitigation budget (line 43).
Strict constraints are used to generate the mitigation path if the potential hazards are captured by the long-term hazard prediction model (with a larger mitigation budget), which satisfies the smoothness and safety constraints during the mitigation process (lines 36-40). In contrast, if a hazard is not predicted by the long-term model but by the short-term model, loose constraints (e.g., wider ranges of $\alpha^j_i$ in Eq. \ref{eq:limit2}) will be applied to ensure fast mitigation within the short mitigation deadline (lines 31-34). These prediction models and mitigation path generation strategies compensate for each other to improve the mitigation performance. 
Then a sequence of corrective control actions $U^c$ is derived by feeding the mitigation path to the mitigation action generation model (\textit{ActNet}) (line 47). 
{Patients are warned about potential safety risks (line 48) and offered a detailed mitigation plan (line 49), which can be used for manual correction or automated mitigation based on their preference.}

\begin{figure*}[t]
    \vspace{1em}
    \centering
    \begin{minipage}{0.64\textwidth}
	\scriptsize \centering
    \includegraphics[width=\textwidth]{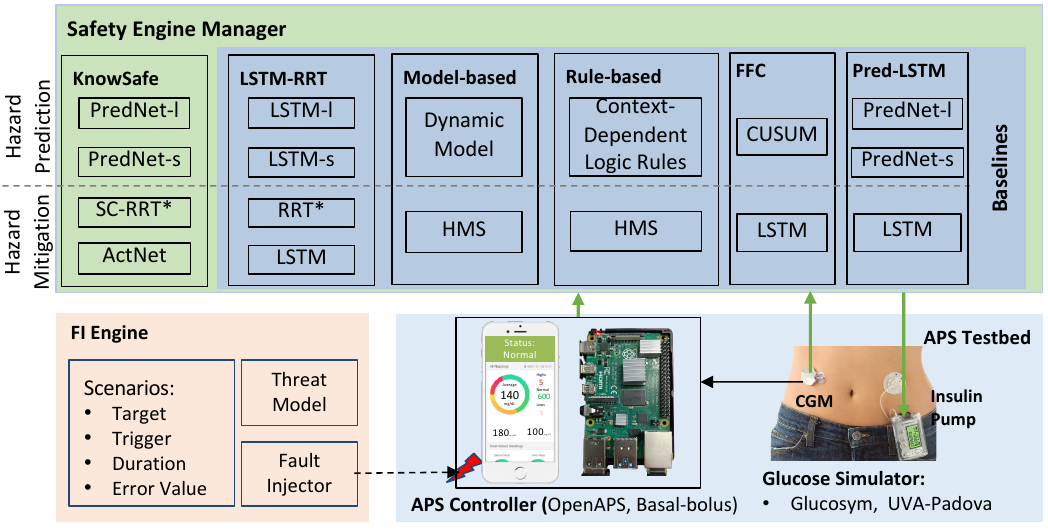} 
    \end{minipage}
    \begin{minipage}{0.35\textwidth}
    \resizebox{\textwidth}{!}{%
    
    \begin{tabular}{|l|l|l|l|l|l|l|}
    \multicolumn{7}{c}{
    \normalsize
    \textbf{Ablation Study and Knowledge Integration Overview.}} \\
\hline
\textbf{\begin{tabular}[c]{@{}l@{}}Safety\\ Engine\end{tabular}} & \textbf{KnowSafe} & \textbf{\begin{tabular}[c]{@{}l@{}}LSTM-\\ RRT\end{tabular}} & \textbf{\begin{tabular}[c]{@{}l@{}}Model\\ Based\end{tabular}} & \textbf{\begin{tabular}[c]{@{}l@{}}Rule\\ Based\end{tabular}} & \textbf{FFC} & \textbf{\begin{tabular}[c]{@{}l@{}}Pred-\\ LSTM\end{tabular}} \\ \hline
\begin{tabular}[c]{@{}l@{}}Short-Term\\ Hazard\\ Prediction\end{tabular} & $\checkmark$ & x & \multirow{2}{*}{x} & \multirow{2}{*}{$\checkmark$} & \multirow{2}{*}{x} & $\checkmark$ \\ \cline{1-3} \cline{7-7} 
\begin{tabular}[c]{@{}l@{}}Long-Term\\ Hazard\\ Prediction\end{tabular} & $\checkmark$ & x &  &  &  & $\checkmark$ \\ \hline
\begin{tabular}[c]{@{}l@{}}Mit. Path \\ Planning\end{tabular} & $\checkmark$ & x & \multirow{2}{*}{$\checkmark$} & \multirow{2}{*}{$\checkmark$} & \multirow{2}{*}{x} & \multirow{2}{*}{x} \\ \cline{1-3}
\begin{tabular}[c]{@{}l@{}}Mit. Action \\ Generation\end{tabular} & $\checkmark$ & x &  &  &  &  \\ \hline

\multicolumn{4}{l}{$\checkmark$: with domain knowledge} \\
\multicolumn{4}{l}{x: without domain knowledge} \\

\end{tabular}%
    }
    \end{minipage}

    \vspace{-0.5em}
    \caption{Left: Experimental Platform with different Safety Engines, integrated with an APS controller running on a PC/cellphone or an embedded device (e.g., Raspberry Pi 4), a Glucose Simulator,  and a Software Fault Injection (FI) Engine. Right: Overview of ablation study and knowledge integration across each safety engine.}
    \label{fig:platform}
    \vspace{-1.5em}
\end{figure*}

When running in the \textit{Mitigation} mode (line 2), the $GetAction$ function sequentially selects a mitigation action $u^c_{kMit}$ from the set $U^c$ (line 14), which will be then sent to be executed by the actuators (line 20).
To balance accuracy and computation overhead, the mitigation path and control action sequences are only updated when the error between the actual state $x_t$ and the expected state in the mitigation path $\Bar{X}$ exceeds a preset threshold $\theta$ (line 9). 
The \textit{Mitigation} mode remains active until the system returns to the target region $\mathcal{X}_*$ or the mitigation budget $\mathcal{D}$ is exhausted (lines 3-5). The user (patient) or physician will be warned of any mitigation failures (e.g., no safe path is found (line 45), or the budget is used up without reaching a safe state (line 7)).


\section{Experimental Evaluation}
\label{sec:evaluation}

\subsection{Experimental Setup}

%
To overcome the last design challenge (C3) mentioned in Section \ref{sec:DesignChallenge}, we develop an open-source simulation environment (see Fig. \ref{fig:platform}) that integrates the closed-loop simulation of two example APS controllers with an adverse event simulator to evaluate different safety engines. {We also evaluate the predictive capabilities of the safety engine using two real-world datasets.}
We run the experiments with both APS controllers and simulators on an x86\_64 PC with an Intel Core i9 CPU @ 3.50GHz and 32GB RAM running Linux Ubuntu 20.04 LTS.  We use TensorFlow v.2.5.0 to train our ML models and Scikit-learn v.1.1.1 for data pre-processing and experimental evaluation.

\textbf{Simulation Testbed.} As shown in Fig. \ref{fig:platform}, the APS testbed integrates two widely-used APS controllers (OpenAPS \cite{openSourceOpenAPS} and Basal-Bolus \cite{BBcontroller}) with two different patient glucose simulators, including Glucosym~\cite{openSourceGlucosym} and UVA-Padova Type 1 Diabetes Simulator (T1DS)~\cite{man2014uva}. The Glucosym simulator contains models of 10 actual Type I diabetes patients \cite{KanderianT1}, while the UVA-Padova T1DS features 30 virtual patients, shown to be representative of the Type 1 diabetes mellitus population observed in a clinical trial \cite{visentin2014university}, and has been approved by the FDA for pre-clinical testing of APS \cite{man2014uva}. This closed-loop testbed was also validated using actual data from a clinical trial, and the simulated data was shown to satisfy the requirements of relevance, completeness, accuracy, and balance \cite{hawkins2021guidance} for developing ML models \cite{zhou2022design}.

In each experiment, the patient simulator interacts with the APS controller for 150 iterations (about 12.5 actual hours), with each step/iteration in the simulation representing 5 minutes in the real APS control system. 
Simulations begin with the patient's initial glucose values within the normal range of 70 to 180 $mg/dL$ without additional meals or exercise during the simulation period, mimicking a scenario of the patient at nighttime after eating dinner. 
To account for inter-patient variability, each system version (without a safety engine and with each different safety engine) is evaluated using ten patient profiles in each glucose simulator. 

{
\textbf{Datasets.}
We cannot evaluate the hazard mitigation approach on actual APS as it requires run-time execution in a closed-loop system, posing safety risks for actual patients. Instead, we evaluate the accuracy of the proposed method in predicting BG sequences using two publicly available datasets: PSO3 \cite{pso3-dataset}, collected during a clinical trial involving 45 individuals over six weeks at three clinical centers in the United States and Canada, and the OpenAPS Data Commons \cite{shahid2022large}, contributed by users of open-source automated insulin delivery systems, comprising over 46,070 days of data and more than 10 million CGM data points.
{The other parts of the method (including hazard and deadline prediction) cannot be evaluated using these existing datasets, as they rarely include hazards and adverse event data.}

}
\subsection{Adverse Event Simulation}

We run the closed-loop simulation of a glucose simulator and an APS controller with a source-level fault injection (FI) engine that directly perturbs the values of the controller’s state variables within their acceptable ranges over a random period to simulate the effect of accidental faults or attack scenarios introduced in Section \ref{sec:threatmodel} (Table \ref{tab:attackscenario}).  
We simulate two categories of attacks: 
\begin{itemize}
\vspace{-0.15em}
    \item \textbf{Availability attacks} set the control output to zero (\textit{Min} scenario for the \textit{Rate} variable in Table \ref{tab:scenarios}) or keep the same sensor measurements after the ${k}_{th}$ control cycle for \textit{m} steps $x_{k:k+m}=x_k$ (\textit{Hold} scenario in Table \ref{tab:scenarios}).
    \item \textbf{Integrity attacks} add a bias $b$ to the sensor readings $x_k=x_k + b$ (\textit{Add}, \textit{Subtract}, \textit{Max}, and \textit{Min} scenarios in Table \ref{tab:scenarios}).
\end{itemize}

\begin{table}[b]
\vspace{-1em}
\centering
\caption{Adverse Event Simulation Experiments Per Patient.}
\vspace{-1em}
\label{tab:scenarios}
\resizebox{\columnwidth}{!}
{%
\begin{threeparttable}
\begin{tabular}{|l|l|p{1cm}p{1cm}|l|}
\hline
\multirow{2}{*}{\textbf{\begin{tabular}[c]{@{}l@{}}Target State Variables\end{tabular}}} & \multirow{2}{*}{\textbf{Scenario}} & \multicolumn{2}{c|}{\textbf{Attack Value}} & \multirow{2}{*}{\textbf{\begin{tabular}[c]{@{}l@{}}No.\\ Simulations\end{tabular}}} \\ \cline{3-4}
 &  & \multicolumn{1}{c|}{\textbf{BG}} & \multicolumn{1}{c|}{\textbf{Rate}} &  \\ \hline
\multirow{5}{*}{\begin{tabular}[c]{@{}l@{}}Blood Glucose (BG)/\\ Insulin Output (Rate)\end{tabular}} & Hold & \multicolumn{2}{c|}{Repeat} & 63 \\ \cline{2-5} 
 & Add & \multicolumn{1}{l|}{{[}32,64{]}} & {[}0.5,1{]} & 126 \\ \cline{2-5} 
 & Subtract & \multicolumn{1}{l|}{{[}32,64{]}} & {[}0.5,1{]} & 126 \\ \cline{2-5} 
 & Max & \multicolumn{1}{l|}{175} & 2 & 63 \\ \cline{2-5} 
 & Min & \multicolumn{1}{l|}{80} & 0 & 63 \\ \hline
\end{tabular}%

\end{threeparttable}
}
\end{table}


%
For each FI scenario, the FI engine determines (i) the target state variable, (ii) the trigger condition for activating the fault, (iii) the duration of the fault, and (iv) the error values to be injected.
In this paper, we utilize a random strategy that randomly chooses from several different start times and durations uniformly distributed within the entire simulation period to inject the faults, resulting in 882 FI simulations for each patient (3 start times $\times$ 3 durations $\times$ 7 initial BG values $\times$ 14 attack values, see the last column of Table \ref{tab:scenarios}) and 17,640 simulations for all 20 patients in both simulators.
This translates into 25.52 years of simulation data (17,640 $\times$ 12.5 hours) used for training and testing different safety engines. 
The details of each FI scenario are shown in Table \ref{tab:scenarios}.

\subsection{Adversarial Training}
\label{sec:advtraining}

We use both normal and hazardous data for training hazard prediction models to improve robustness. For evaluation of the trained state and hazard prediction models, we consider the whole simulation trace as a sample and label it as hazardous if any state sequence in that simulation trace overlaps the unsafe region (e.g., BG value higher than 180 mg/dL or less than 70 mg/dL).
%
For the training and testing of response control action generation models, we use a sequence of collected control actions $u_{t:t+n}$ as the ground truth values. The input to the models is the sequence of expected or target system states $x_{t+1:t+n}$ upon sequential execution of the control actions.

Our FI engine results in 3,230 (38.3\%) simulations with hazards in the Glucosym simulator and 3,315 (37.6\%) simulations with hazards in the UVA-Padova simulator, which are used for training and testing each ML model for hazard prediction and mitigation.
We use a 4-fold cross-validation setup for training and testing each patient-specific ML model. Specifically, for each patient, we train separate ML models for state prediction and mitigation action generation using data from 75\% of the simulation traces and test the models on remaining 25\% simulation traces.

To reduce the model complexity and computational resources, we minimize the number of neural network layers. After exploring various architectures, the best-performing model was a two-layer (128-64 units) stacked LSTM. 



    
\subsection{Metrics}
We introduce the following metrics to evaluate the performance of the proposed methodology. 
\begin{itemize}[leftmargin=*]
\item \textbf{\textit{Prediction Accuracy}} 
is measured using the root mean squared error (RMSE) between the predicted/generated trajectories and the ground truth trajectories. 
We also utilize the standard binary metrics of false positive rate (FPR), false negative rate (FNR), and F1 score to evaluate the accuracy of hazard prediction and mitigation action activation. 
We consider the whole trajectory of each simulation as a sample for evaluating performance. 

\item \textbf{\textit{Reaction Time}} measures the timeliness of hazard prediction and indicates the maximum time budget to mitigate and prevent potential hazards.  It is the time difference between when the prediction network detects the hazard and when the system enters a hazardous state. {A larger reaction time indicates a more sufficient mitigation budget and improved hazard prediction timeliness.}

\item \textbf{\textit{Mitigation Path Planning Efficiency}} is measured by evaluating the performance of each algorithm in finding an optimal mitigation path using the following two metrics:
\begin{itemize}[leftmargin=*]
  \item \textbf{\textit{Convergence Rate} }is the percentage of simulations in which a mitigation path ends up in a final state in the safe region.
\item \textbf{\textit{Satisfaction Rate} }is the percentage of simulations with a valid mitigation path that satisfies safety constraints. 
\end{itemize}

\item \textbf{\textit{Mitigation Outcome}} metrics evaluate the overall performance of the safety engine (the hazard prediction and mitigation pipeline):

\begin{itemize}
    \item \textbf{\textit{Mitigation Success Rate (MSR)}} is calculated as the percentage of hazardous simulations in which the system transitions into the unsafe region without mitigation but after implementing the mitigation actions returns to the target region.
    \item \textbf{\textit{Out-of-Safe-Range Rate}} is the percentage of simulations with state variable values falling outside of the safe range (e.g., higher than 180 mg/dL or less than 70 mg/dL in APS). This metric measures the remaining hazard rate even with the mitigation algorithm, the summation of which is the complementary set of the \textit{Mitigation Success Rate}.
    \item \textbf{\textit{Max Deviation}} measures the maximum distance from the safe region boundaries. This metric evaluates the risk of transitioning to an unsafe state with or without safety engine.
    \item \textbf{\textit{New Hazard Rate}} represents the percentage of non-hazardous simulations where false alarms or unnecessary mitigation actions introduce new hazards.

\end{itemize}

\end{itemize}

Final results are reported by calculating  average value of each metric over all cross-validation folds across all patients. 





\subsection{Prediction Accuracy}
\label{sec:Evaluation:accuracy}



\begin{table}[b]
\vspace{-1.5em}
\caption{Performance of Each ML Model in Predicting System States and Estimating Deadlines of Mitigation Averaged over 8,820 Simulations in Glucosym and UVA-Padova Simulators, respectively. RMSE is measured in $mg/dL$ for state estimation and \textit{Iteration} (representing 5 minutes in actual APS) for deadline estimation.}
\vspace{-1em}
\label{tab:acc}
\resizebox{\columnwidth}{!}
{%
\begin{threeparttable}

\begin{tabular}{|c|l|l|lll|l|}
\hline
\multirow{2}{*}{\textbf{Simulator}} & \multicolumn{1}{c|}{\multirow{2}{*}{\textbf{Model}}} & \textbf{State} & \multicolumn{3}{c|}{\textbf{Hazard Prediction}}                                            & \textbf{Deadline} \\ \cline{4-6} 
                                    & \multicolumn{1}{c|}{}                                & \textbf{RMSE}  & \multicolumn{1}{l|}{\textbf{FNR}}   & \multicolumn{1}{l|}{\textbf{FPR}}   & \textbf{F1}    & \textbf{RMSE}     \\ \hline
\multirow{4}{*}{Glucosym}           & LSTM-l                                               & 2.34          & \multicolumn{1}{l|}{0.037}          & \multicolumn{1}{l|}{0.091}          & 0.909          & 0.89             \\ \cline{2-7} 
                                    & LSTM-s                                               & 0.61          & \multicolumn{1}{l|}{0.008}          & \multicolumn{1}{l|}{0.017}          & 0.982          & 0.17            \\ \cline{2-7} 
                                    & PredNet-l                                            & \textbf{1.46} & \multicolumn{1}{l|}{\textbf{0.014}} & \multicolumn{1}{l|}{\textbf{0.061}} & \textbf{0.943} & \textbf{0.64}    \\ \cline{2-7} 
                                    & PredNet-s                                            & \textbf{0.38} & \multicolumn{1}{l|}{\textbf{0}}     & \multicolumn{1}{l|}{\textbf{0.003}} & \textbf{0.997} & \textbf{0.13}    \\ \hline
\multirow{4}{*}{UVA-Padova}         & LSTM-l                                               & 7.66          & \multicolumn{1}{l|}{0.213}          & \multicolumn{1}{l|}{0.091}          & 0.899          & 2.24             \\ \cline{2-7} 
                                    & LSTM-s                                               & 3.14          & \multicolumn{1}{l|}{0.045}          & \multicolumn{1}{l|}{0.082}          & 0.951          & 0.21             \\ \cline{2-7} 
                                    & PredNet-l                                            & \textbf{7.20} & \multicolumn{1}{l|}{\textbf{0.015}} & \multicolumn{1}{l|}{\textbf{0.039}} & \textbf{0.961} & \textbf{2.07}    \\ \cline{2-7} 
                                    & PredNet-s                                            & \textbf{2.85} & \multicolumn{1}{l|}{\textbf{0}}     & \multicolumn{1}{l|}{\textbf{0.012}} & \textbf{0.993} & \textbf{0.20}    \\ \hline
\end{tabular}%

\end{threeparttable}

}
\end{table}

Table \ref{tab:acc} presents the performance of the proposed prediction models ({PredNet}) for system state and hazard prediction as well as mitigation deadline estimation in both glucose simulators compared to the baseline long short-term memory (LSTM) models that share the same architecture with the {PredNet} {but without knowledge integration}. We use "l" and "s" to differentiate the ML models with long-term and short-term prediction windows (e.g., 24 and 6 control cycles in this paper), respectively.
The final results are reported by calculating the average value of each metric over all cross-validation folds across all patients. 

With simulation data from Glucosym, we see that ML models trained under the guidance of domain knowledge achieve the best prediction accuracy with up to 37.7\% lower root mean squared error (RMSE) (0.38 vs. 0.61) in state prediction and 28.1\% lower RMSE (0.64 vs. 0.89) in deadline estimation.  
Although the {PredNet}s only has slightly better F1 scores in hazard prediction (1.5\%-3.8\% improvement), they achieve zero false negative rates (FNR), which is important for hazard prediction and mitigation, while keeping false positive rate (FPR) up to 82.3\%  (0.003 vs. 0.017) lower.
(The above binary metrics are evaluated by considering the whole trajectory of each simulation as a sample.)

For tests in the UVA-Padova simulator, we observe similar performance improvements of combined knowledge and data driven ML models in hazard prediction (4.4\%-6.9\%) and system state prediction and deadline estimation, indicating \textit{the advantage of combining knowledge with data-driven approaches in predicting hazards and stable performance across different simulators and patients}.
However, the overall prediction error is higher in the UVA-Padova simulator than in the Glucosym simulator for all the models. This might be because the UVA-Padova simulator also includes a noise model for sensor measurement \cite{Simglucose}.

\begin{figure}[t]
    \centering
     \includegraphics[width=\columnwidth]{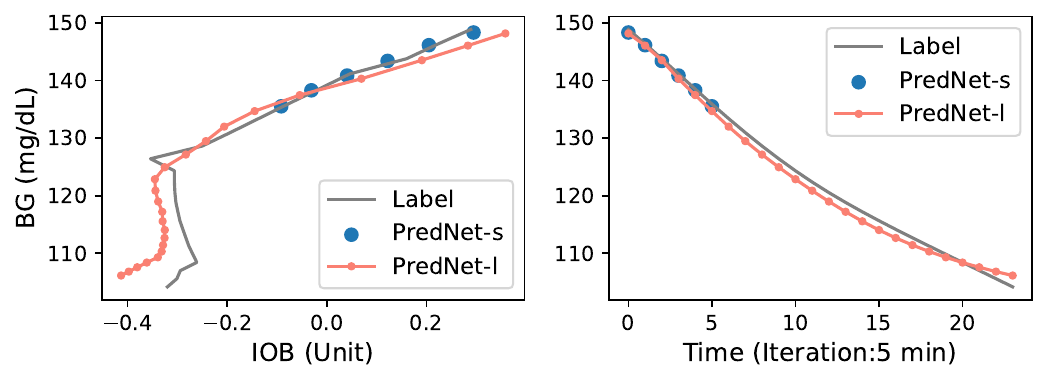}
    
    \vspace{-1em}
    \caption{Example State Trajectories Predicted by Both Long-term and Short-term Hazard Prediction Models.}
    \label{fig:PredictNet}
\end{figure}

In Table \ref{tab:acc}, we also observe PredNet-s achieve up to 74.0\% (0.38 vs. 1.46) and 90.3\% (0.20 vs. 2.07) reduction of RMSE in state and deadline prediction, respectively, compared to PredNet-l,  with at least 69.2\% lower FPR (0.012 vs. 0.039) in predicting the occurrence of hazards while keeping a zero negative rate. 
This further attests to the benefit of employing two-level hazard prediction models, leveraging both the high accuracy of PredNet-s models to not miss any positive cases and the extended prediction window of PredNet-l models to preemptively mitigate potential hazards effectively. 
An example of the state trajectory and the predictions of PredNets is also shown in Fig. \ref{fig:PredictNet}. As we can see, the PredNet-s predictions almost overlap with the ground-truth state trajectory, and the long-term predictions by PredNet-l also approximate the ground truth well.

\begin{table}[t]
    \centering
    \caption{Prediction Error on Existing Datasets.}
    \vspace{-0.5em}
    \begin{tabular}{|l|c|c|} 
    \hline
     \textbf{Model}    & PSO3 \cite{pso3-dataset} & OpenAPS Data Commons \cite{shahid2022large}\\ \hline
     ODE-based \cite{Dutta2018} &7.19 mg/dL & N/A\\ \hline
      PredNet   & 6.35 mg/dL& 7.69 mg/dL\\ \hline
    \end{tabular}
    \label{tab:rmse-dataset}
\end{table}

To further evaluate the performance of PredNet in actual APS and reduce the sim-to-real gap, we also train and evaluate the combined knowledge and data driven ML models for system state sequence prediction on two publicly available clinical trial datasets, PSO3 \cite{pso3-dataset} {and OpenAPS Data Commons \cite{shahid2022large}}. 
As shown in Table \ref{tab:rmse-dataset}, our PredNet can achieve a slightly smaller RMSE than the ML models developed in a previous work (6.349 vs. 7.187 $mg/dL$ on average across the same patient data) \cite{Dutta2018} using PSO3 and a similar RMSE of 7.691 mg/dL on OpenAPS dataset. 

\subsection{Evaluation of Mitigation Path Generation}


\begin{table}[t]
\footnotesize
\centering
\caption{Performance of Each Mitigation Path Generation Algorithm (Averaged over All the Simulations). }
\label{tab:resRRT}
\resizebox{\columnwidth}{!}
{%
\begin{threeparttable}

\begin{tabular}{|l|l|l|l|l|l|}
\hline
\multirow{2}{*}{\textbf{Simulator}} & \multirow{2}{*}{\textbf{\begin{tabular}[c]{@{}l@{}}Prediction \\ Model\end{tabular}}} & \multirow{2}{*}{\textbf{Algorithm}} & \multirow{2}{*}{\textbf{\begin{tabular}[c]{@{}l@{}}Conv.\\ Trials\end{tabular}}} & \multirow{2}{*}{\textbf{\begin{tabular}[c]{@{}l@{}}Path \\ Len\end{tabular}}} &
\multirow{2}{*}{\textbf{\begin{tabular}[c]{@{}l@{}}Satisfaction\\ Rate\end{tabular}}} \\
 &  &  &  &  &   \\ \hline

\multirow{4}{*}{Glucosym}   & \multirow{2}{*}{PredNet-s}  & SC-RRT*     & 174 & 15  & 89.4\% \\ \cline{3-6} 
                            &                     & RRT* & 32  & 10  & 6.4\%  \\ \cline{2-6} 
                            & \multirow{2}{*}{PredNet-l} & SC-RRT*     & 194 & 25 &  85.8\% \\ \cline{3-6} 
                            &                     & RRT* & 41  & 14 &  3.4\%  \\ \hline
\multirow{4}{*}{UVA-Padova} & \multirow{2}{*}{PredNet-s}  & SC-RRT*     & 245 & 10 & 93.5\% \\ \cline{3-6} 
                            &                     & RRT* & 78  & 8  &  31.9\% \\ \cline{2-6} 
                            & \multirow{2}{*}{PredNet-l} & SC-RRT*     & 167 & 22 & 94.7\% \\ \cline{3-6} 
                            &                     & RRT* & 50  & 17 &  32.0\% \\ \hline

\end{tabular}%

\begin{tablenotes}
\item[*] A unit of path length represents 5 minutes in actual APS. \textit{Conv. Trials}: the number of trials each RRT algorithm takes to converge to a valid path.
\end{tablenotes}

\end{threeparttable}

}
\end{table}

Table \ref{tab:resRRT} presents the results of the mitigation path generation algorithm, SC-RRT*, in comparison with a baseline RRT* algorithm that does not consider application-based constraints (shown in Eq. \ref{eq:limit1}-\ref{eq:limit2} and the example in Table \ref{table:appconstraints}).  

\begin{table}[t]
\begin{center}
\caption{Domain-Specific Constraints for APS. (BG is in \textit{mg/dL} and IOB is in \textit{unit}).}
\vspace{-0.75em}
\label{table:appconstraints}
\resizebox{\columnwidth}{!}
{%
    
\begin{tabular}{ | c | l |l| c | l |l|} \hline
    \textbf{No.} & \textbf{Description} &\textbf{Constraint}&\textbf{No.} & \textbf{Description} &\textbf{Constraint} \\ \hline
    1 & $ dBG/dt $ & [-5, 3] & 3 & $ dIOB/dt $ & [-0.1, 0.1]\\ \hline
    2 & $ d^2BG/dt^2 $ & [-2.5, 2.5] & 4 & $ d^2IOB/dt^2 $ & [-0.05, 0.05]\\ \hline
\end{tabular}
}

\end{center}
\vspace{-2em}
\end{table}
Both algorithms converge to a solution for all the experiments in our test. However, the baseline algorithm achieves a much lower satisfaction rate (the last column of Table \ref{tab:resRRT}, calculated by the percentage of simulations with a valid mitigation path that satisfies the safety constraints), as it does not check the physical constraints or safety requirements during the path generation process. 

\begin{figure}[t]
    \centering
    \includegraphics[width=\columnwidth]{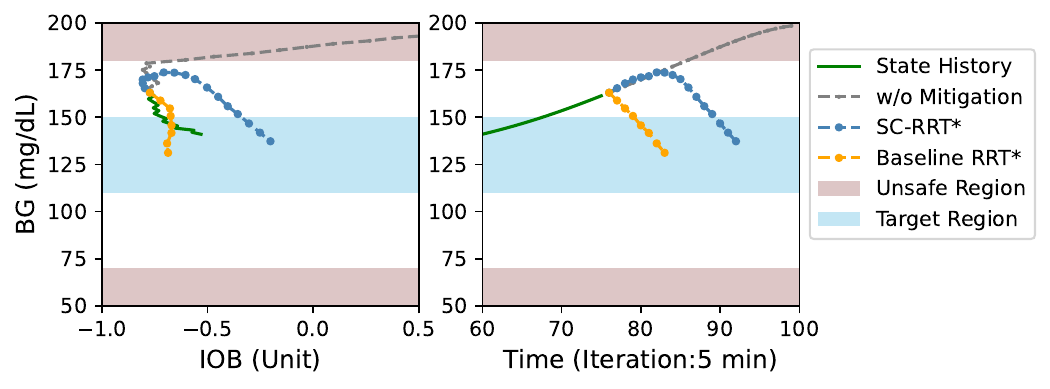}
    \vspace{-1.5em}
    \caption{Example Mitigation Paths by SC-RRT* and Baseline RRT* Algorithms in the State Space (Left) and Time Space (Right).}
    \label{fig:rrtPath}
    \vspace{-0.5em}
\end{figure}

An example of the mitigation path generated by both algorithms is shown in Fig. \ref{fig:rrtPath}.
We observe that the baseline RRT* (marked by orange dots) generates a sharp turn at the early mitigation stage, which is either impossible to achieve due to the considerable lag in the biological BG process, or 
may cause patient discomfort due to the sharp changes in the BG level. On the other hand, the proposed SC-RRT* algorithm (marked by blue dots) generates a smoother trajectory for hazard mitigation by considering the rate and trends of changes as well as physiological constraints. 
Although the proposed SC-RRT* algorithm takes more trials to converge to an optimal path, the average time is much smaller than the length of a control cycle and, thus, will not affect the system operation. We further assess the time overhead of each model in Section \ref{sec:utilization}.

In Table \ref{tab:resRRT}, we also see that the mitigation paths generated based on hazards inferred by the {PredNet-s} model have shorter path lengths (averaged over all the patients in each simulator) than those generated based on the {PredNet-l} predictions. 
This observation attests to the efficiency of the proposed PredNet-s in ensuring a valid mitigation path within a short mitigation deadline as complementary to the PredNet-l that mainly ensures smoothness in state transitions and user experience.

\subsection{Evaluation of Response Action Generation}

\begin{figure}[t]
        
    \centering
    \includegraphics[width=0.6\columnwidth]{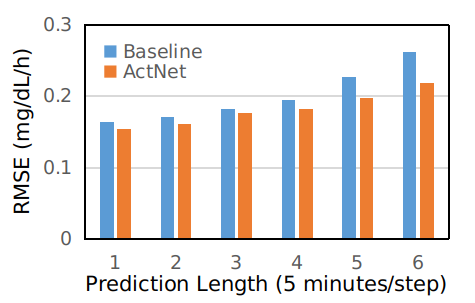}
    \vspace{-0.5em}
    \caption{Performance of Control Action Generation.}

    \label{fig:actionseq}
    \vspace{-1em}
\end{figure}

Fig. \ref{fig:actionseq} presents the performance of the response action generation models with and without the integration of HMS knowledge in accurately reconstructing the control action sequence and replicating the controller dynamics.
We see that the ML model trained using the proposed custom loss function (Eq. \ref{eq:loss_custom}) (ActNet) maintains a lower RMSE than the baseline LSTM model for all the prediction lengths, indicating \textit{the advantage of combining knowledge and data in accurately generating mitigation/recovery actions}.

\begin{table*}[ht!]
\vspace{1em}
\centering
\caption{Mitigation Performance and Time Overhead of KnowSafe vs. Baselines. 
The severity of hypoglycemia (BG<70 $mg/dL$) increases with decreasing BG value, including
mild (54 $mg/dL$<BG<70 $mg/dL$) and more severe (BG<54 $mg/dL$) categories.
}
\vspace{-0.5em}
\label{tab:mitigationres}
{%

\begin{tabular}{|l|l|lll|l||l|l|}
\hline
\multirow{2}{*}{\textbf{Method}} & \multirow{2}{*}{\textbf{No. Sim.}} & \multicolumn{3}{c|}{\textbf{Out-of-Safe-Range Rate}} & \multirow{2}{*}{\begin{tabular}[c]{@{}l@{}}\textbf{MSR (Mitigation}\\ \textbf{Success Rate)}\end{tabular}} & \multicolumn{2}{c|}{\textbf{Time Overhead (ms)}}\\ \cline{3-5} \cline{7-8}

            &      & \multicolumn{1}{l|}{\textbf{BG\textgreater{}180 $mg/dL$}} & \multicolumn{1}{l|}{\textbf{BG\textless{}70 $mg/dL$}} & \textbf{BG\textless{}54 $mg/dL$} &      & Normal Mode &  Mit. Mode    \\ \hline
KnowSafe    & 6,545 & \multicolumn{1}{l|}{\textbf{428 (6.5\%)}}         & \multicolumn{1}{l|}{\textbf{44 (0.7\%)}}      & \textbf{0}               & \textbf{92.8\%} & 46.5 & 54 or 0\\ \hline
LSTM-RRT    & 6,545 & \multicolumn{1}{l|}{2492 (38.1\%)}                & \multicolumn{1}{l|}{603 (9.2\%)}              & 66 (1\%)                 & 52.7\%      & 46.2 &  81.5 or 0  \\ \hline
Rule-based  & 6,545 & \multicolumn{1}{l|}{1421 (21.7\%)}                & \multicolumn{1}{l|}{1792 (27.4\%)}            & 370 (5.7\%)              & 50.9\%       & 0.1 (106.5 $\mu$s) &  0.1 (121.5 $\mu$s) \\ \hline
Model-based & 6,545 & \multicolumn{1}{l|}{1481 (22.6\%)}                & \multicolumn{1}{l|}{3685 (56.3\%)}            & 716 (10.9\%)             & 21.1\%        & 87.9 & 87.9  \\ \hline
FFC         & 6,545 & \multicolumn{1}{l|}{499 (7.6\%)}                  & \multicolumn{1}{l|}{459 (7.0\%)}              & 39 (0.6\%)               & 85.4\%        & 28.8  & 28.8   \\ \hline
Pred-LSTM         & 6,545 & \multicolumn{1}{l|}{467 (7.1\%)}                  & \multicolumn{1}{l|}{285 (4.4\%)}              & \textbf{0}                & 88.5\%     & 46.5 &  28.7   \\ \hline
\end{tabular}%
}
\vspace{-1em}
\end{table*}

We also observe that the RMSE of the ActNet is slightly lower than that of the baseline for the one-step prediction. However, with the increase in the prediction length, the difference between the ActNet and baseline increases as well. This is because the integrated domain knowledge is independent of the data and prediction length.
This observation demonstrates \textit{more obvious benefit of integrating domain knowledge on context-specific mitigation actions with ML models for improving sequential prediction accuracy than single value regression, maintaining a stable and low RMSE, and reducing the performance degradation}. 

Furthermore, the integrated domain knowledge (HMS) helps improve the explainability of black-box ML models in safety-critical CPS, as it offers a reason for selecting a corrective control action under a given context and can be used for run-time verification of the ML model outputs. 

\subsection{End-to-End Safety Engine Evaluation}
\label{sec:Evaluation:pipeline}

We integrate the proposed safety engine (the pipeline of {PredNet}, {SC-RRT*}, {ActNet} models) as well as several state-of-the-art baselines with the controllers in our closed-loop testbed and compare their performances. 

\subsubsection{Baselines}
\label{sec:baseine}
We {conduct ablation studies to} compare our combined knowledge and data-driven approach, \textit{KnowSafe}, with solely knowledge-driven and data-driven approaches by designing different baselines, including a rule-based approach that uses the same domain knowledge on constraints and mitigation actions as \textit{KnowSafe}, a model-based approach, and ML-based approaches that share the same architecture (see Fig. \ref{fig:platform}).

The \textbf{rule-based} baseline includes (i) an anomaly detection algorithm designed based on the context-dependent specification of unsafe control actions and (ii) a mitigation algorithm developed based on the generated HMS in Table \ref{table:stltable}. The detection rules specify whether a control action issued by the controller under a given context might result in transition into the unsafe region.~
The mitigation rules specify the corrective actions to be taken in each context to stay within the safe region.

We re-implement state-of-the-art defense solutions (e.g., CI \cite{choi_ccs_18} or SAVIOR \cite{SAVIOR2020}) as a \textbf{model-based} baseline using a dynamic model of the physical system (patient physiology), called the Bergman \& Sherwin model~\cite{KanderianT1}, to estimate the possible BG value ($BG_{t+1}$) after executing the pump's command ($u_t$) on the patient's current state ($BG_t$). If the predicted BG value goes beyond the patient's normal range ([70,180] mg/dL, as defined by the medical guidelines), a mitigation algorithm similar to the rule-based mitigation will issue a corrective action that replaces the control action by the controller until the system state returns to the target range.

We also design two solely data-driven baseline models. 
The first baseline is developed by integrating LSTM models and RRT* with the same architecture as \textit{KnowSafe} but without the integration of domain-specific knowledge (referred to as \textbf{LSTM-RRT}). For the LSTM models, this is equivalent to setting $\eta$ to be 1 in Eq. \ref{eq:loss_reachablestate} and Eq. \ref{eq:loss_custom}.
The second data-driven model directly maps the sensor measurement to the control output $y_{ML}$ with a single regression model, created based on a feed-forward control (\textbf{FFC}) method proposed by \cite{pidpiper2021} for unmanned vehicles. In this approach, when the accumulated error between the ML output and the controller output exceeds a preset threshold, the mitigation mode is activated, and the controller output is replaced by the ML output to be delivered to the actuator. 
(Note that since these previous works are not directly transferable to APS, we re-implement their high-level pipeline but customize their models and parameters.)

To evaluate the effectiveness of the proposed hazard mitigation module, we design another baseline, \textbf{Pred-LSTM}, that has the same prediction module (PredNet) as \textit{KnowSafe} but utilizes a single LSTM model (with HMS integration) to generate corrective control actions (without path planning part). The details of each baseline are shown in Fig. \ref{fig:platform}.

\subsubsection{Mitigation Outcome}
We rerun all the hazardous simulations (6,545 in both simulators) with different safety engines (\textit{KnowSafe} vs. baselines) and present their performance in mitigating hazards and maintaining the system inside the safe region in Table \ref{tab:mitigationres}.

We see that the rule-based baseline mitigates 50.9\% of the hazards
, but fails to keep the remaining half of the simulations within the safe region. Similarly, the LSTM-RRT baseline successfully prevents 52.7\% of the hazards but has 3,095 simulations outside the safe region.
This is because while the rule-based strategy knows about the context, the generated high-level mitigation action does not infer specific quantitative values for mitigating potential hazards.
In contrast, the ML baseline outputs more specific mitigation actions but relies on a black-box data-driven model that may violate HMS rules under a given context.
Therefore, \textit{by combining the knowledge with data, KnowSafe overcomes the shortage of either approach and utilizes the advantages of both methods, achieving at least 41.9\% higher success rate than solely rule-based and ML-based (LSTM-RRT) baselines and keeping the number of out-of-safe-range simulations low.}



We also observe that \textit{KnowSafe} outperforms the model-based and FFC baselines by achieving the highest MSR and the fewest simulations with BG outside the target range. In addition, BG goes below 54 $mg/dL$ in 39 and 716 simulations with FFC and model-based mitigation, respectively, which may result in severe hypoglycemia and serious complications (e.g., seizure, coma, or even death). In contrast, \textit{KnowSafe} keeps the BG above 54 $mg/dL$ for all the simulations. Although BG goes below 70 $mg/dL$ in 44 simulations with \textit{KnowSafe}, the percentage is much lower than the baselines and is much less concerning than the cases with BG below 54 $mg/dL$.

The better performance of the proposed mitigation approach over these baselines might be because \textit{our approach aims to be early in hazard prediction and mitigation by estimating the possibility of the system state entering the unsafe region}. In contrast, approaches like the FFC and model-based baselines detect the hazards \textit{after} they occur or wait until the error between the ML predictions and the actual system states or outputs exceeds a noticeable threshold, which wastes the limited mitigation time budget on hazard detection and may lead to mitigation/recovery failure.

Although Pred-LSTM has the same prediction module with \textit{KnowSafe} and shares the same benefit of early hazard prediction, its overall MSR is 4.3\% lower. This result demonstrates the advantage of our hazard mitigation pipeline (SC-RRT* and ActNet) over a single ML model that directly outputs a control action based on current system states since SC-RRT* can find an optimal mitigation path while Pred-LSTM can only generate a single-step mitigation action at each control cycle. Since Pred-LSTM has the same prediction module as \textit{KnowSafe}, we don't compare its performance in the following sections.

\begin{table}[t]
\vspace{1em}
\centering
\caption{Average BG Value and Max Deviations in the Adverse Event Simulations with No Mitigation or with KnowSafe.}
\vspace{-1em}
\label{tab:deviation}
\resizebox{\columnwidth}{!}
{%
\begin{tabular}{|l|ll|l|}
\hline
\multirow{2}{*}{\textbf{Metric}} &
  \multicolumn{2}{l|}{\textbf{Max Deviation ($mg/dL$)}} &
  \multirow{2}{*}{\textbf{\begin{tabular}[c]{@{}l@{}}Average BG\\ ($mg/dL$)\end{tabular}}} \\ \cline{2-3}
              & \multicolumn{1}{l|}{\textbf{Above 180}} & \textbf{Below 70} &       \\ \hline
Simulations w/o Mitigation & \multicolumn{1}{l|}{175.3}              & 49.6              & 97.3  \\ \hline
Simulations with KnowSafe & \multicolumn{1}{l|}{21.6}               & 4.3               & 128.9 \\ \hline
\end{tabular}%
}
\vspace{-1.5em}
\end{table}
Table \ref{tab:deviation} also shows the maximum deviations of BG from the safe region boundaries in each simulation with or without the proposed safety engine. The maximum deviation from the lower bound of the safe region is reduced from 49.6 $mg/dL$ to 4.3 $mg/dL$ (91.3\% lower). Similarly, the maximum deviation from the upper bound of the safe region is reduced from 175.3 $mg/dL$ to 21.6 $mg/dL$ (87.6\% lower). Therefore, \textit{KnowSafe} maintains a BG level between the range of [65.7, 201.6] with an average value of 128.9 $mg/dL$ inside the target range, successfully preventing all the hypoglycemia and hyperglycemia hazards. 

\subsubsection{Timeliness}

\begin{figure}[t]
    \centering
    \includegraphics[width=\columnwidth]{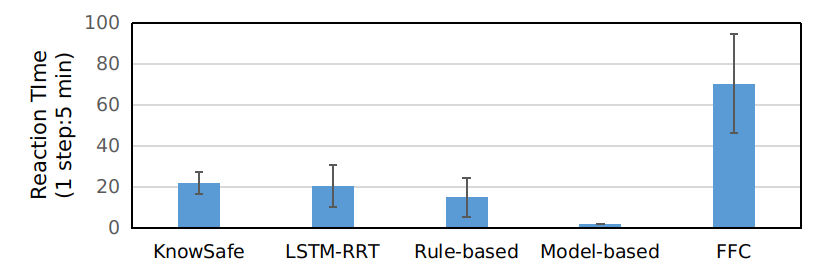}
    \vspace{-2em}
    \caption{Reaction Time of Each Mitigation Strategy (a larger value is better).}
    \label{fig:reactiontime}
    \vspace{-1.5em}
\end{figure}

Fig. \ref{fig:reactiontime} presents the timeliness of each mitigation strategy in predicting potential hazards measured by the average reaction time. We have the following observations:

\begin{itemize}[leftmargin=*]
    \item The model-based baseline keeps the lowest reaction time and MSR (see Table \ref{tab:mitigationres}), indicating that sufficient reaction time is necessary for effective mitigation/recovery.
    \item The LSTM-RRT baseline achieves a similar reaction time as \textit{KnowSafe} and 1.4 times higher MSR than the model-based approach, further attesting to the benefit of early hazard prediction. However, the LSTM-RRT baseline still needs to be revised to mitigate the remaining 49.1\% hazards, indicating that solely being early is not enough to guarantee mitigation success. 
    \item The rule-based baseline has a smaller reaction time than the LSTM-RRT approach but achieves a similar MSR, indicating that being context-aware improves mitigation performance. 
    \item The FFC baseline achieves the largest reaction time by generating alerts across the entire simulation period, which would bother the users and may result in unnecessary mitigation. We will further analyze such false positive rates in the following subsection.
    \item \textit{KnowSafe} maintains a stable and sufficient reaction time and achieves the highest MSR, demonstrating \textit{the advantage of combining domain-specific knowledge with data in being both context-aware and early in the prediction of hazards for successful mitigation/recovery}.
\end{itemize}

\begin{figure}[t]
    \centering
    \includegraphics[width=\columnwidth]{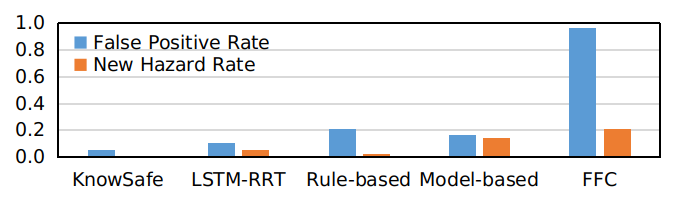}
    \vspace{-2.8em} 
    \caption{Performance of Different Safety Engines in 11,095 Non-Hazardous Simulations.}
    \label{fig:mitigation-FP}
    \vspace{-1.5em}
\end{figure}

\subsubsection{False Positives}
We also evaluate the consequences of the unwanted mitigation actions activated by false alarms of each mitigation strategy by running it with all the non-hazardous simulations (11,095 in both simulators). 
We see in Fig. \ref{fig:mitigation-FP} that the FFC baseline is falsely activated for 96.9\% simulations and results in 21.3\% new hazards, which indicates the drawbacks of such a single ML model in accurately detecting and mitigating hazards.
The LSTM-RRT baseline achieves a lower FPR and new hazard rate than the model-based and FFC baselines, indicating the benefit of the proposed hazard prediction and mitigation pipeline. 
The rule-based safety engine gets two times of false positives than the LSTM-RRT one but keeps the new hazard rate much lower, demonstrating the low risk of the generated HMS even with false activation.  
\textit{KnowSafe} maintains the lowest FPR (5.4\%) without introducing any new hazards, attesting to \textit{the advantage of combining domain knowledge with data-driven techniques in reducing the FPR and the risk of introducing new hazards in the hazard prediction and mitigation process.} 


\subsection{Mitigation of Stealthy Attacks}
To evaluate the performance of the proposed mitigation strategy against stealthy attacks, we assume a powerful attacker who knows the mitigation mechanism, including the ML architecture and parameters. We rerun the experiments shown in Table \ref{tab:scenarios}, in which the attacker stops the attack action (or injection of faults) to stay stealthy when such action will trigger the mitigation alarms if launched. We compare the performance with a model-based baseline (similar to existing solutions CI and SAVIOR) and an ML-based baseline (e.g., LSTM-RRT introduced in Section \ref{sec:baseine}). We do not test the FFC baseline due to its high FPR. 
\begin{table}[b]
\caption{Performance of Each Mitigation Strategy against Stealthy Attacks.}
\vspace{-1em}
\label{tab:stealthyresult}
\resizebox{\columnwidth}{!}
{%
\begin{tabular}{|l|l|ll|}
\hline
\multirow{2}{*}{\textbf{Method}} & \multirow{2}{*}{\textbf{\begin{tabular}[c]{@{}l@{}}No. Simulation Out of\\ Safe Range [70,180] mg/dL\end{tabular}}} & \multicolumn{2}{l|}{\textbf{Max Deviation (mg/dL)}} \\ \cline{3-4} 
 &  & \multicolumn{1}{l|}{\textbf{Above 180}} & \textbf{Below 70} \\ \hline
KnowSafe & 0 & \multicolumn{1}{l|}{0} & 0 \\ \hline
LSTM-RRT & 2,958 & \multicolumn{1}{l|}{45.3} & 15.9 \\ \hline
Model-based & 5,026 & \multicolumn{1}{l|}{74.6} & 26.5 \\ \hline
\end{tabular}%
}
\end{table}

In Table \ref{tab:stealthyresult}, we see that the success rate of \textit{KnowSafe} under stealthy attacks is 100\%, and BG values are kept within the safe range [70,180] $mg/dL$ in all the simulations with both the Glucosym simulator and the UVA-Padova simulator. 
In comparison, the model-based approach fails to mitigate the stealthy attack in 5,026 simulations (76.8\% of 6,545 hazardous cases in Table \ref{tab:mitigationres}) with a max deviation of 74.6 $mg/dL$ above 180 $mg/dL$ and 26.5 $mg/dL$  below 70 $mg/dL$.  
This is because the model-based approach can only predict the system state in the next time step, which is not sufficient to prevent or mitigate hazards under stealthy attacks. 
On the other hand, the ML-based approach LSTM-RRT reduces max deviations to 45.3 $mg/dL$ above 180 $mg/dL$ and 15.9 $mg/dL$ below 70 $mg/dL$ but still fails to mitigate 2,958 (45.2\%) stealthy attacks due to unconstrained predictions.

In contrast, the proposed \textit{KnowSafe} overcomes these problems by combining knowledge and data that can accurately predict system states in the following two hours, which thus ensures enough reaction time and a high success rate in mitigating stealthy attacks.

{
\subsection{Mitigation of Sensor Attacks}
\label{sec:sensorattack}
Although this paper focuses on mitigating attacks or faults targeting APS controllers and relies on existing solutions to detect and mitigate sensor attacks, some sensor attacks may evade detection by current anomaly detection methods and affect the performance of our proposed approach. To evaluate this, we implement a cumulative sum (CUSUM)-based method \cite{attack_pcs} to detect sensor attacks before the data is sent to KnowSafe. Specifically, we train an ML model to predict the expected system state $\hat{x}_{t}$ based on historical system states $X_{t-1}$ and control actions $U_{t-1}$:

\begin{equation}
    \hat{x}_{t} = MLmodel.predict(X_{t-1},U_{t-1})
\end{equation}

An alert is raised if the accumulated error $S(t)$ in sensor measurements exceeds a threshold $\tau$ while considering a bias parameter $b$ to avoid false positives during normal operations (without attacks).  

\begin{equation}
    S(t) = max(0,S(t-1) + \delta - b); \delta = |{x}_{t} - \hat{x}_{t} |
\end{equation}

We train the model on our simulation dataset, both with and without fault injections, using 80\% of the data for training and 20\% for testing. Experimental results show that the implemented method detects all simulated attacks when the threshold $\tau$ is set below a certain value (see Fig. \ref{fig:roc}).

\begin{figure}[b]
    \vspace{-2em}
    \begin{minipage}{0.48\columnwidth}
	\small 
        \centering
    \includegraphics[width=\linewidth]{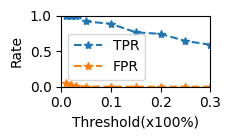}
    \end{minipage}
    \begin{minipage}{0.5\columnwidth}
        \centering
        \resizebox{\columnwidth}{!}{%
    \begin{tabular}{|ll|l|l|}
\hline
\multicolumn{2}{|l|}{\textbf{Sensor Compromised?}} & \textbf{Yes} & \textbf{No}\\ \hline

\multicolumn{2}{|l|}{\textbf{No. BG>180 mg/dL}} & 7.8\% & 6.5\%\\ \hline
\multicolumn{2}{|l|}{\textbf{No. BG<70 mg/dL}} & 3.5\% & 0.7\%\\ \hline
\multicolumn{2}{|l|}{\textbf{Miti. Success Rate}} & 88.7\% & 92.8\%\\ \hline
\multicolumn{1}{|l|}{\multirow{2}{*}{\begin{tabular}[c]{@{}l@{}}\textbf{Max BG}\\ \textbf{Deviations}\end{tabular}}} & \textgreater{}180 mg/dL & 22.8 & 21.6 \\ \cline{2-4} 
\multicolumn{1}{|l|}{} & \textless{}70 mg/dL & 6.3 & 4.3 \\ \hline
\multicolumn{2}{|l|}{\textbf{Average BG} (mg/dL)} & 127.1 & 128.9\\ \hline
\end{tabular}%
}
    \end{minipage}

    \vspace{-1em}
    \caption{Performance of Sensor Attack Detection (Left) and KnowSafe Mitigation with Detection Threshold Set at 3\% (Right). }
    \label{fig:roc}
\end{figure}

For stealthy attacks, we model a surge attack, which aims to maximize damage to the system as quickly as possible, and a bias attack, which attempts to modify the system discretely by adding small perturbations over an extended period of time \cite{attack_pcs,biehler2024sage}. We observe that the surge attack can only be sustained for a few control cycles to avoid detection, limiting its impact on deviating the system’s states. Similarly, the small perturbations introduced by the bias attack do not significantly alter the control actions enough to result in unsafe states.


To further evaluate the impact of the simulated sensor attacks on our mitigation performance when the patient does not take any action in response to the raised alerts, we adopt a checkpointing protocol \cite{Checkpoint2018Kong} and reconstruct the system states based on trustworthy historical sensor data, which KnowSafe then uses to generate mitigation actions.
Experimental results in Fig. \ref{fig:roc} show a 4.1\% decrease in overall MSR due to errors in state reconstruction. Additionally, the detection latency introduced by the anomaly detection method reduces the available time for mitigation. Nevertheless, KnowSafe maintains comparable effectiveness in mitigating hazards (see Table \ref{tab:mitigationres}).
A more in-depth study of advanced sensor attacks is beyond the scope of this paper.



}

\subsection{Resource Utilization}
\label{sec:utilization}
We evaluate the resource utilization of the proposed approaches in the closed-loop simulation. 
We run the simulations with different safety engines or without any mitigation/recovery 1,000 times and calculate the average time overhead. Results in Table \ref{tab:mitigationres} show that \textit{KnowSafe} introduces an average overhead of 46.5 ms (PredNet) during normal operation and before entering mitigation mode and 54 ms (SC-RRT* 25.3 ms, ActNet 28.7 ms) when in the mitigation mode. These overheads are much smaller than the control cycle period in APS (5 minutes), and thus will not severely affect the system operation at run-time. 
Note that \textit{KnowSafe} does not introduce any time overhead after entering the mitigation mode when no updates to the corrective actions are made (see lines 9-13 in Algorithm \ref{alg:mitigation}).
The LSTM-RRT baseline has a similar time overhead for hazard prediction and mitigation but a slightly larger overhead for the baseline RRT* algorithm (52.8 ms) as more nodes are found and processed due to looser constraints. 
In contrast, the overall time overhead of rule-based, model-based, and FFC baselines are 121.5 $\mu$s, 87.9 ms, and 28.8 ms, respectively.
This shows that the overhead of \textit{KnowSafe} is comparable to single models used in model-based and FFC baselines.

We also evaluate the resource utilization of \textit{KnowSafe} on an actual typical MCU used in APS (Raspberry Pi 4 with 8GB RAM as shown in Fig. \ref{fig:platform}) with optimized ML implementation \cite{TensorFlowLite}. We observe a time overhead of 18 $\mu$s ($4.12\times10^{-7}$\% of the normal execution time of 4.36s) and 25.3 ms ($5.8\times10^{-5}$\%) before and after entering the mitigation mode, respectively, and the peak memory usage increase is 38MB (0.48\% of the available 8GB RAM). {We maintain a simple model architecture to meet the strict resource requirements of edge devices, making it easier to verify as well.}


\section{Discussion}
\label{sec:discussion}
\textbf{Sensor Perturbations:}
This paper mainly focuses on accidental faults or malicious attacks targeting the CPS controller. Any perturbations in the sensor data will potentially affect the safety engine's performance. However, benefiting from the knowledge of data and model properties, the proposed PredNet can construct a sequence of system states with a low RMSE (e.g., 0.38 $mg/dL$ in APS), which can be used to detect sensor compromise (similar to \cite{Choi2020software, SAVIOR2020}) and generate mitigation/recovery actions for a considerable period. Previous work has also shown that integrating domain knowledge with ML anomaly detection can improve the robustness against small perturbations on ML inputs without sacrificing accuracy and transparency \cite{zhou2022robustness}. 

\textbf{Data Impact:}
{
The performance degradation in predicting unseen data and scenarios or corner cases is a common problem for data-driven methods. We try to address this problem by integrating domain knowledge with data \cite{zhou2022robustness} and using adversarial training \cite{chen2025revisiting}. 
We include the most common types of attack scenarios reported in the literature and real databases for adversarial training.

}

\textbf{Sim-to-Real Gap:}
The differences between the simulation and the actual implementation in the real world might also threaten the validity of the proposed method.
We try to reduce the sim-to-real gap by using real control software (used with actual diabetic patients \cite{lewis2020yourself,lewis2016real}), real patient simulators that are either approved by the FDA for clinical testing or use actual patient profiles \cite{man2014uva, KanderianT1}, and a realistic adverse event simulator.  
Furthermore, two publicly available datasets collected from a clinical trial \cite{pso3-dataset} and diabetes community \cite{shahid2022large} and an actual implementation on a typical APS MCU are used to further attest to the effectiveness/validity of the proposed approach.

\textbf{Generalization and Limitation:} We demonstrate the effectiveness of the proposed approach in two closed-loop APS testbeds and two real-world datasets, but several limitations narrow the generalization of our approach. 
Specifically, our approach is limited to the CPS in which the system state can be represented by a subset of state variables, based on which the {(un)safe region} and safety properties can be formally specified. Also, our approach is limited to CPS, where the control cycle and device memory are larger than this approach's requirement reported in Section \ref{sec:utilization}.

The primary task of adapting our method to a new type of CPS involves creating safety specifications, which may not be easy for more complex systems and might require domain-specific technical expertise. This paper adopts a formal framework that can generate formal safety specifications using a control-theoretic hazard analysis method \cite{dsn2021zhou,zhou2023hybrid} in cooperation with domain experts. An example of safety specification for Adaptive Cruise Control systems (ACC) for autonomous driving was explored in previous work \cite{zhou2023hybrid}.
Further study of the knowledge specification and hazard mitigation in different systems and scenarios is beyond the scope of the paper.


\textbf{Real-World Deployment:}
The proposed approach can be implemented on a smartphone (similar to some commercial APS as shown in Fig. \ref{fig:platform}) or be integrated with the pump microcontroller, but running ML models on resource-scarce embedded systems might be challenging. 
However, significant progress has been made in implementing ML models on embedded systems, such as using low-power embedded GPUs \cite{microchipGPU}), ML accelerators \cite{ISM330DHCX}, or optimized ML implementations \cite{gupta2017protonn,TensorFlowLite}.
Further, some high-end MCUs have large memory and powerful computation capabilities \cite{Armv8}. 
In this paper, we also implemented the proposed method on a typical MCU used in APS (Raspberry Pi 4 with 8GB RAM as shown in Fig. \ref{fig:platform}) with improved inference time (see Section \ref{sec:utilization}).
In addition, the integration of domain knowledge with the ML models is done offline during training without adding computational costs at run-time.

Our models are trained on patient/configuration-specific data. Changes in configuration, patient’s physiological, or environmental context necessitate recalibration/retraining. 

\vspace{-0.5em}
\section{Related work}

\textbf{Anomaly Detection and Attack Recovery:} 
In previous sections, we introduced existing works on anomaly detection and recovery using model-based, rule-based, or ML-based approaches. 
Our work distinguishes itself from these previous works in combining knowledge with data-driven techniques for preemptive hazard prediction and mitigation with better accuracy and a higher mitigation success rate. 
Further, this paper focuses on mitigating or recovering the CPS from accidental faults or malicious attacks that directly compromise the controller software or hardware functionality in addition to detecting sensor attacks. 

\textbf{Security of APS:} 
We reviewed existing APS security work in Section \ref{sec:DesignChallenge}. To the best of our knowledge, this paper is the first on run-time hazard/attack mitigation in APS.

\textbf{Knowledge Integration:} 
Integrating expert knowledge into ML has been an active area of research.
Previous works have focused on enforcing ML to follow logic rules during the training process through applying soft constraints \cite{xu2018semantic}, designing specific logistic circuits \cite{liang2019learning} and network structures \cite{chen2019embedding}, generating graph models \cite{guo2016jointly}, or utilizing knowledge distillation \cite{gou2021knowledge}.
In this paper, we integrate domain-specific knowledge of safety constraints and properties as soft constraints with the multivariate sequential prediction models for anomaly detection and hazard mitigation.

\section{Conclusion}

This paper proposes a combined knowledge and data driven approach for the design of safety engines that can predict and mitigate hazards in APS. We integrate expert domain knowledge with machine learning through the design of novel custom loss functions that enforce the satisfaction of domain-specific safety constraints during the training process. Experimental results for two closed-loop APS testbeds and two diabetic datasets show that the proposed safety engines achieve increased accuracy in predicting hazards and a higher success rate in mitigating them. 
Future work will focus on studying the generalization of the proposed approach to other CPS. %

\section*{Acknowledgment}

This work was supported in part by the Commonwealth of Virginia under Grant CoVA CCI: CQ122-WM-02 and by the National Science Foundation (NSF) under grants CNS-2146295 and CCF-2402941.

\bibliographystyle{IEEEtran}
\bibliography{main}

\begin{IEEEbiography}[{\includegraphics[width=1in,height=1.25in,clip,keepaspectratio]{{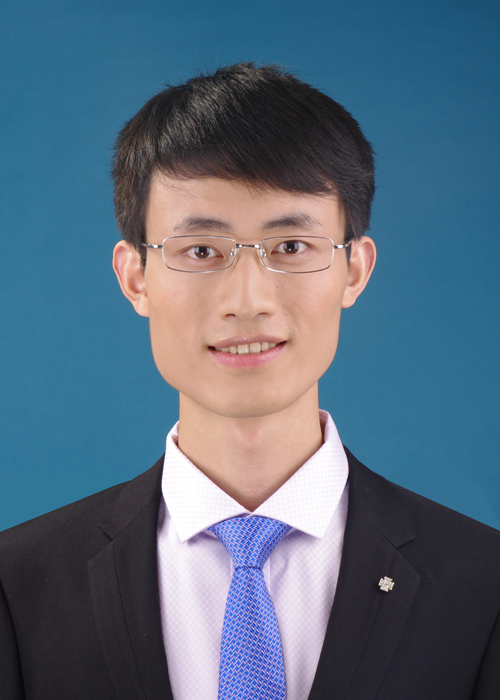}}}]%
{Xugui Zhou}
obtained his B.E. in Automation and M.E. in Control Science and Engineering from Shandong University, China, and his Ph.D. in Electrical and Computer Engineering from the University of Virginia. He is currently an assistant professor in the Department of Electrical and Computer Engineering, with a joint appointment in the Department of Computer Science, at Louisiana State University. Before pursuing his Ph.D., he was an engineer at NR Research Institute, State Grid of China, and he researched power grid protection technology. His research interests are at the intersection of computer systems dependability and control system theory and engineering by drawing techniques from formal methods and machine learning. 
He is the recipient of the NSF Cyber-Physical Systems Rising Star Award and is the inventor of three international patents.
\end{IEEEbiography}

\vspace{-1em}

\begin{IEEEbiography}[{\includegraphics[width=1in,height=1.25in,clip,keepaspectratio]{{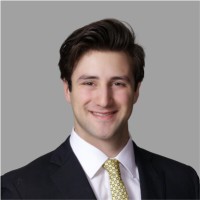}}}]%
{Maxfield Kouzel} received his B.Sc. in Computer Science from the University of Virginia. His research interests center around building more effective and robust autonomous systems with control theory and artificial intelligence.
\end{IEEEbiography}

\vspace{-1em}
\begin{IEEEbiography}[{\includegraphics[width=1in,height=1.25in,clip,keepaspectratio]{{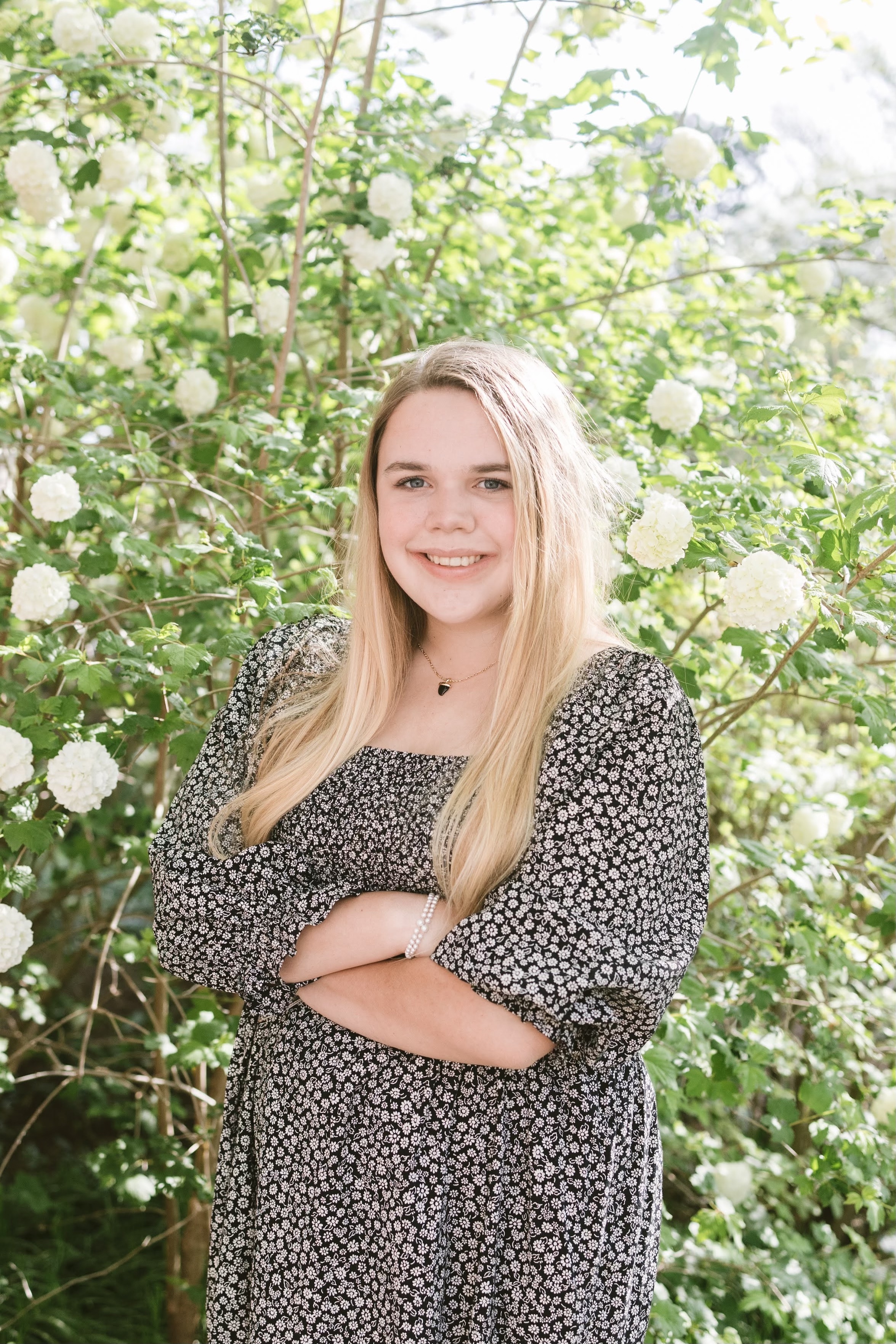}}}]%
{Chloe Smith} received her B.Sc. and M.Sc. in Computer Science from the University of Virginia. Her research interests center around improving the safety and user experience of artificial pancreas systems.
\end{IEEEbiography}

\vspace{-1em}
\begin{IEEEbiography}[{\includegraphics[width=1in,height=1.25in,clip,keepaspectratio]{{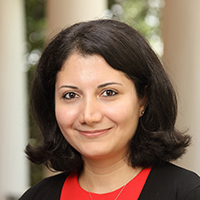}}}]%
{Homa Alemzadeh} received her B.Sc. and M.Sc. in Computer Engineering from the University of Tehran and her Ph.D. in Electrical and Computer Engineering from the University of Illinois Urbana-Champaign. 
She is currently an associate professor in the Department of Electrical and Computer Engineering, with a courtesy appointment in the Department of Computer Science at the University of Virginia. She is also affiliated with the LinkLab, a multi-disciplinary center for research and education in Cyber-Physical Systems (CPS). 
Before joining UVA, she was a research staff member at the IBM T. J. Watson Research Center.
Her research interests are at the intersection of computer systems dependability and data science. 
She is particularly interested in data-driven resilience assessment and design of embedded and cyber-physical systems, with a focus on safety and security validation and monitoring in medical devices and systems, surgical robots, and autonomous systems. She is the recipient of the 2022 NSF CAREER Award and the 2017 William C. Carter Ph.D. Dissertation Award in Dependability from the IEEE TC and IFIP Working Group 10.4 on Dependable Computing and Fault Tolerance. 
\end{IEEEbiography}

\end{document}